\documentclass[journal,twocolumn]{IEEEtran}   %ÔËÐÐ1ÏÂ
\usepackage{graphicx}
\usepackage{epstopdf}
\usepackage{amssymb}
\usepackage{amsfonts}
\usepackage{amsmath}
\usepackage{algorithm}
\usepackage{algorithmic}
\usepackage{subeqnarray}
\usepackage{cases}
\usepackage{bm}
\usepackage{subfigure,amsmath,amssymb,cite}
\usepackage{stfloats}

\usepackage{color}
\usepackage{float}
\usepackage{array}
\UseRawInputEncoding
\ifCLASSINFOpdf
\else
\fi
\begin{document}

\title{Beamforming Design for IRS-and-UAV-Aided Two-Way Amplify-and-Forward Relay Networks in Maritime IoT}

\author{Xuehui~Wang,~Feng Shu,~Yuanyuan~Wu,~Weiping~Shi,~Shihao~Yan,~Yifan~Zhao,~Qiankun~Cheng,~Zhongwen~Sun~  and~Jiangzhou~Wang,~\emph{Fellow},~\emph{IEEE}

\thanks{This work was supported in part by the National Natural Science Foundation of China (Nos.U22A2002, and 62071234), the Hainan Province Science and Technology Special Fund (ZDKJ2021022), the Scientific Research Fund Project of Hainan University under Grant KYQD(ZR)-21008, and the Collaborative Innovation Center of Information Technology, Hainan University (XTCX2022XXC07). \emph{(Corresponding authors: Feng Shu, Yuanyuan Wu)}.}

\thanks{Xuehui~Wang,~Yuanyuan~Wu, Yifan~Zhao, Qiankun~Cheng, and Zhongwen~Sun are with the School of Information and Communication Engineering, Hainan University,~Haikou,~570228, China.}

\thanks{Feng Shu is with the School of Information and Communication Engineering and Collaborative Innovation Center of Information Technology, Hainan University, Haikou 570228, China, and also with the School of Electronic and Optical Engineering, Nanjing University of Science and Technology, Nanjing 210094, China (e-mail: shufeng0101@163.com).}

\thanks{Weiping Shi is with the School of Network and Communication, Nanjing Vocational College of Information Technology, Nanjing 210023, China.}

\thanks{Shihao Yan is with the School of Science and Security Research Institute, Edith Cowan University, Perth, WA 6027, Australia (e-mail:
s.yan@ecu.edu.au).}

\thanks{Jiangzhou Wang is with the School of Engineering, University of Kent, Canterbury CT2 7NT, U.K. (e-mail: j.z.wang@kent.ac.uk).}

}

\maketitle
\begin{abstract}

In this paper, an intelligent reflecting surface (IRS)-and-unmanned aerial vehicle (UAV)-assisted two-way amplify-and-forward (AF) relay network in maritime Internet of Things (IoT) is proposed, where ship1 ($\text{S}_1$) and ship2 ($\text{S}_2$) can be viewed as data collecting centers. To enhance the message exchange rate between $\text{S}_1$ and $\text{S}_2$, a problem of maximizing minimum rate is cast, where the variables, namely AF relay beamforming matrix and IRS phase shifts of two time slots, need to be optimized.
To achieve a maximum rate, a low-complexity alternately iterative (AI) scheme based on zero forcing and successive convex approximation (LC-ZF-SCA) algorithm is presented.
To obtain a significant rate enhancement, a high-performance AI method based on one step, semidefinite programming and penalty SCA (ONS-SDP-PSCA) is proposed.
\textcolor{blue}{Simulation results show that by the proposed LC-ZF-SCA and ONS-SDP-PSCA methods, the rate of the IRS-and-UAV-assisted AF relay network  surpass those of with random phase and only AF relay networks. Moreover, ONS-SDP-PSCA perform better than LC-ZF-SCA in aspect of rate.}

\end{abstract}


\section{Introduction}

\textcolor{blue}{With the advancement of information technologies in maritime Internet of Things (IoT), the maritime devices such as ships, buoys, offshore platforms and sensor nodes have experienced sustained growth \cite{NKG2023}.} Meanwhile, with the promotion of the Belt and Road strategy, the marine economy has flourished, so that marine activities, such as marine tourism, marine transportation, maritime rescue and marine scientific research, have more stringent requirements on rate and reliability \cite{Wt2021}.
The well-known international marine satellite (i.e., Inmarsat) system can provide some communication services, such as fax, telegraph, and telephone, for the remote areas far away from the coast. However, it is subject to higher communication cost and lower data rate compared to  terrestrial 5G network \cite{Lxl2020,Zjh2019,Zjh2021}.
Aiming to enhance data rate of marine IoT, some high-throughput satellites, such as the Inmarsat-5 satellite network and the Iridium NEXT system, have been launched \cite{Zq2017,MO2021}. Nevertheless, there simultaneously exists a large amount of communication delay. Moreover, the marine electromagnetic propagation environment caused by different weather conditions is extremely complex, which leads to low reliability in satellite-based communication system \cite{KM1969}. In addition, the shore-based communication system suffers from limited coverage and blind zones. Therefore, it is of great significance to build an innovative marine IoT with low cost, high throughput, high transmission efficiency, and extended coverage.

Owing to a set of features of low cost, autonomy, mobility, high flexibility \textcolor{blue}{and} existence of line-of-sight links, unmanned aerial vehicle (UAV) has been popularly applied to terrestrial wireless network. With the ability to communicate and process signal, UAV can be regarded as a aerial base station (BS) or relay node to assist the key data collection and dissemination, so that high-rate and reliable transmission can be obtained \cite{Jsy2020,Glh2020,Mxd2021}. Different from terrestrial wireless network, the communication environment is unstable and the vessels are sparsely distributed in the maritime scenario. However, due to a variety of advantages of UAV mentioned above, some existing research work has emerged where UAV can be regarded as a aerial platform integrated to maritime communication network (MCN) for coverage enhancement.
A reliable maritime machine-type communication (MTC) is necessary for maritime IoT systems, in order to respond to the challenges faced by MTC, namely energy consumption and efficiency, the authors in \cite{RP2020} proposed a novel UAV-aided wireless communication network and applied a genetic algorithm based on probabilistic to solve a maximizing handover efficiency problem.
To provide a high-quality service for more maritime users in remote zones, \textcolor{blue}{an} UAV-assisted MCN based on non-orthogonal multiple access was developed in \cite{Tr2021}, where the intragroup power allocation and the transmission durations among the UAVs were iteratively optimized to maximize the minimum throughput.
%The optimization of UAV trajectory plays an important role for the information collection rate of all USVs in maritime UAV-based IoT systems, to efficiently promote data collection rate of USVs, the authors in \cite{Ll2022} designed a fast UAV trajectory planning method where the problem of optimizing UAV trajectory was converted to that of vehicle route for solution by using the Fermat-point theory.
Although UAV can bring many conveniences to MCN because of unique characteristics, its size, power supply and weight are the constraints, which results in many challenges to UAV-enabled MCN, such as high-complexity power optimization and high-rate data transmission.

\textcolor{blue}{Intelligent reflecting surface (IRS) is a man-made plane, which consists of lots of passive and low-cost reflecting elements \cite{Wxh2022}.} Due to the fact that IRS is able to intelligently regulate the propagation path of reflected signal without power consumption \cite{Dlm2020_1}, \textcolor{blue}{IRS has been considered as a potential solution to meet the high-performance transmission requirement of wireless network \cite{Wqq2019}.} By flexibly deploying IRS on the surface of various buildings and tuning the reflection coefficient (i.e., amplitude and phase) of each unit \cite{Sf2022_Jxy}, the uncontrollability of conventional wireless environment can be broken out, so that the signal coverage can be extended and the capacity of wireless network can be improved \cite{Dlm2020_2}. Because of a set of exclusive advantages, IRS has received increasing research mention from academia and industry.
\textcolor{blue}{So far, IRS has been well studied in several UAV-aided terrestrial communication scenarios \cite{Pxw2021,Alj2021,MA2021,Mh2021,Pxw2022}. The authors introduced the combination of IRS and UAV to air-ground networks to boost its throughout in \cite{Pxw2021}, where two cases of IRS deployment were considered, that is, IRS was mounted on a mobile UAV and IRS was installed on a building. The results proved that the combination could provide better communication service for air-ground networks.} The capacity of a flying IRS-aided UAV wireless network was derived in \cite{Alj2021}, where IRS elements were with certain phases before reflecting signal to UAV. An IRS radio network based on UAV was presented in \cite{Mh2021}, where IRS was used to reflect signal transmitted by UAV to BS, so that UAV transmission can be improved. While meeting minimum master rate demand, a scheme based on relaxation was put forward to design IRS coefficient matrix, IRS scheduling and UAV trajectory for minimum bit error rate.
Except for the terrestrial scenarios mentioned above, IRS has been further researched in maritime scenario. The authors investigated an IRS-aided MCN to provide an effective coverage with low cost in \cite{Zzy2020}, where the maximum effective sum rate can be obtained by jointly optimizing the service time of each ship and IRS beamforming at coastal BS and ship.

%However, the rate gain achieved by IRS-aided UAV wireless network is limit. It is difficult to
%employing only passive IRSs like [15] is difficult to support high-speed communication services, especially under long-distance transmission,

In contrast to IRS, the traditional relay can also forward signal by using amplifying \cite{BS2009}, decoding \cite{BR2006}, compressing \cite{YE2010} and coding \cite{LK2008} strategies in uncontrollable maritime communication environment, so that the received signal can be significantly enhanced to improve communication quality of maritime IoT.
Aiming to improve the reliability of channel, multiple ships were regarded as collaborative relays and introduced to the distributed MCNs \cite{Cxy2022}. To further obtain maximum energy efficiency, a problem of resource allocation was formulated and solved through an iterative optimization algorithm.
In \cite{Dry2020}, a cooperative multicast problem was investigated in a relay-aided maritime wireless network. By alternatively designing the beamforming vectors at BS and processing matrices at relays, the total transmit power can be minimized under the constraints of signal-to-interference-plus-noise ratio.
Although the traditional relay has significant advantages in signal processing, it is an active forwarder which needs more costly hardware and circuit, higher energy and power consumption to improve rate performance compared to IRS \cite{FG2009,Qg2021}.
%In \cite{Fh2022}, the researchers compared performance difference between the two wireless networks respectively assisted by IRS and decode-and-forward (DF) relay. It was demonstrated that the rate of the DF-aided system is higher than that of the IRS-aided network when the transmitter and receiver are far apart. However, the IRS-aided network is superior to the DF-aided system when the goal is to harvest maximum energy efficiency with minimum transmit power.
At present, network coverage is still one of the most important and fundamental capabilities of maritime communication systems.
%The deployment and operation costs of BSs in the fifth-generation (5G) mobile network have further increased compared to the fourth-generation (4G) network, and the electric charge of 5G BSs has brought a significant burden to operators.
Consequently, it is urgent and imperative to develop an innovative, efficient, high-rate, low-cost, and low-power-consumption solution for MCN.

Given that the advantages of IRS and relay, the combined network of IRS and relay is extremely attractive, which is considered as a win-win strategy in perspective of cost, spectrum and energy efficiency improvement, coverage extension and rate performance enhancement. With increasing attention paid to the hybrid network of relay and IRS, some corresponding work has emerged at present. \cite{Sq2021,MZT2022,Zbx2021,Wxh2023_AF,Cxy2023}.
In \cite{Sq2021}, multiple IRSs were applied to reflect signals in the decode-and-forward (DF) relay network, and the ergodic capacity was derived and analysed.
%Compare with a single IRS-aided DF relay system, the multi-IRS-assisted DF relay network could obtain rate gain by optimizing the numbers of IRSs and IRS units.
A novel wireless network assisted by an IRS was proposed in \cite{Zbx2021}, where the IRS controller was acted as a DF relay. By optimizing the time allocations of two time slots for the DF relay and passive beamforming at IRS, the coverage range and rate performance of the proposed novel network could be significantly improved.
In \cite{Cxy2023}, the authors proposed to integrate the combination network of DF relay and IRS into MCN. By designing the work mode of each ship (i.e., DF relay and IRS or IRS), transmit power of DF relay, transmit beamforming vector of BS and phase shifts of IRS, the total transmit power of the MCN can be minimum.

To our best knowledge, most of the research work on the combination network is focused on the DF relay, the research work on the hybrid network of amplify-and-forward (AF) relay and IRS is little, especially in marine communication scenarios.
Additionally, IRS is fixed on tall buildings in most IRS-assisted terrestrial wireless networks, which leads to limitations and inflexibility in the deployment of IRS.
In particular, when the direct link from IRS to transceiver is obstructed, the rate performance and the coverage can be greatly affected, which results in poor communication for edge users.
Motivated by this, considering the unique characteristics of IRS, AF relay and UAV, an IRS-and-UAV-aided two-way AF relay network in maritime IoT is proposed, where IRS is mounted to UAV and two ships are considered as data centers responsible for collecting a large amount of data from buoys, offshore platforms and sensor nodes. With the aid of IRS and relay, the two ships can exchange their information to realize data interoperability. The contributions are summarized as follow:

\begin{enumerate}

\item

%A maximizing minimum rate optimization problem is established, where the transmit power of AF relay is limited and each IRS phase shift is need to meet the requirement of unit-modulus.
%Aiming to achieve a high information exchange rate, a low-complexity algorithm based on zero forcing and successive convex approximation (LC-ZF-SCA) is put forward to address the optimization problem by alternately optimizing one variable and fixing the other two variables. For AF relay beamforming matrix, its semi-closed expression can be derived by utilizing ZF method. For the optimization subproblem of IRS phase shift for the first or second time slots, SCA algorithm is applied to solve the non-convex subproblem given the other two variables. In contrast to random phase and only AF relay, the proposed LC-ZF-SCA method can achieve a 68.5\% higher rate gain when total transmit power is 30dBm.

\textcolor{blue}{An optimization problem of maximizing minimum rate is cast, which is subject to the transmit power of AF relay and unit-modulus requirement of IRS phase shift. With the aim of solving the problem, a low-complexity alternating iterative optimizing (AIO) algorithm based on zero forcing and successive convex approximation (LC-ZF-SCA) is came up with. We use ZF method to harvest the closed-form expression of AF relay beamforming matrix, and utilize SCA algorithm to address the non-convex optimization subproblems of IRS phase shifts of two time slots. Furthermore, the highest order of the computational complexity of LC-ZF-SCA method is $\mathcal{O}(N^3+M^3)$ FLOPs. In contrast to the comparison benchmarks like random phase and only AF relay, up to 68.5\% rate gain is attained by the proposed LC-ZF-SCA method when total transmit power is 30dBm.}

\item

\textcolor{blue}{With the goal of rate enhancement, a high-performance AIO algorithm based on one step, semidefinite programming and penalty SCA (ONS-SDP-PSCA) is proposed. For AF relay beamforming matrix, singular value decomposition (SVD) and ONS method are applied to derive its closed-form solution. Afterwards, the optimization problem is reformulated as a SDP problem. Firstly, the non-convex subproblem corresponding to IRS phase shift matrix in the second time slot is transformed into convex by generalized fractional programming (GFP). Then the subproblems of IRS phase shift matrices of two time slots are tackled via PSCA algorithm. Accordingly, its highest order of the computational complexity is $\mathcal{O}(N^{6.5}+M^2)$ FLOPs. From the simulation result, the proposed ONS-SDP-PSCA method obtains higher rate than LC-ZF-SCA method especially in the high total transmit power region.}

\end{enumerate}

The rest of this article is arranged as follows. In Section \ref{SystemModel}, we propose an IRS-and-UAV-assisted two-way AF relay network in maritime IoT, construct its system model and formulate optimization problem. In Section \ref{LC-ZF-SCA}, a low-complexity method is put forward to solve the optimization problem. In Section \ref{ONS-SDP-PSCA}, a high-performance scheme is proposed to improve rate. The related numerical results are analyzed in Section \ref{Results}, and conclusions are shown in Section \ref{Conclusions}.

 \emph{Notations}: The letters of lower case, bold lower case, and bold upper case are used to denote scalars, vectors and matrices. The conjugate, transpose, conjugate transpose, Moore-Penrose pseudo inverse and trace of a matrix are respectively represented by $(\cdot)^*$, $(\cdot)^T$, $(\cdot)^H$, $(\cdot)^\dag$ and $\text{tr}\{\cdot\}$. The expectation operation, absolute value, 2-norm \textcolor{blue}{and Frobenius norm} are respectively denoted as $\mathbb{E}\{\cdot\}$, $|\cdot|$,  $\|\cdot\|$ \textcolor{blue}{and $\|\cdot\|_F$}. $\arg(\cdot)$ and $\mathfrak{R}(\cdot)$ stand for the phase and real part of a complex number, respectively. In addition, $\mathbf{I}_M$ is a $M \times M$ identity matrix.

\section{System Model and Problem Formulation} \label{SystemModel}
\subsection{Signal Model}
\begin{figure}[htb]
\centering
\includegraphics[width=0.480\textwidth]{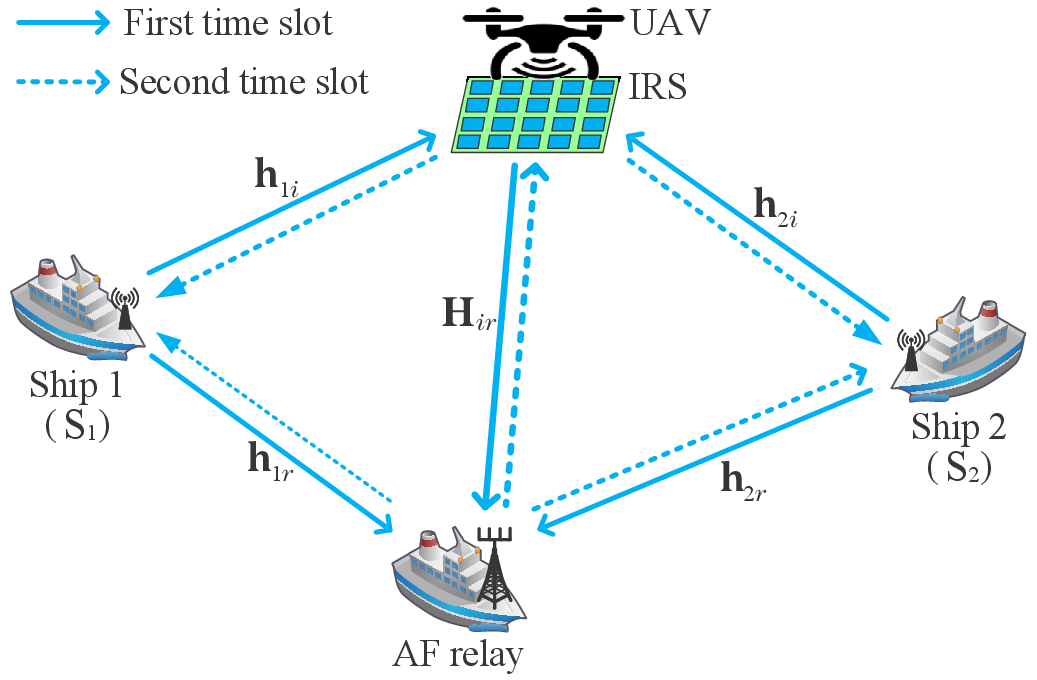}
\centering
\caption{An IRS-and-UAV-assisted two-way AF relay network in maritime IoT.}
\label{System_Model}
\end{figure}
Fig. \ref{System_Model} sketches an IRS-and-UAV-aided two-way AF relay network in maritime IoT, where two single-antenna ships can be regarded as data collecting centers. Via an IRS attached to UAV and an AF relay, ship1 ($\text{S}_1$) and ship2 ($\text{S}_2$) can exchange their information collected from maritime IoT devices like buoys and sensor nodes for data interoperability.
The AF relay is equipped with $M$ antennas, while the IRS is made up of $N$ passive reflecting elements. It is assumed that the UAV hovers and keeps static at a desired position, the direct link between two ships is obstructed, and global \textcolor{blue}{channel state informations (CSIs)} can be perfectly known.

In the first time slot, S1 and S2 simultaneously transmit their signals to AF relay. The received signal at AF relay can be denoted as
\begin{align}\label{yr}
\mathbf{y}_{r}&= \sqrt{P_1}(\mathbf{h}_{1r}+\mathbf{H}_{ir}\bm{\Theta}_1\mathbf{h}_{1i})x_1  \nonumber\\
&\quad + \sqrt{P_2}(\mathbf{h}_{2r}+\mathbf{H}_{ir}\bm{\Theta}_1\mathbf{h}_{2i})x_2 + \mathbf{n}_r,
\end{align}
where $x_1$ and $x_2$ are the independent signal from $\text{S}_1$ and $\text{S}_2$, and $\mathbb{E}\{x_1^H x_1\}=\mathbb{E}\{x_2^H x_2\}=1$. $P_1$ and $P_2$ respectively denote the transmit power of $\text{S}_1$ and $\text{S}_2$. Without loss of generality, it is assumed that all channels follow Rayleigh fading. $\mathbf{h}_{1r}\in  \mathbb{C}^{M \times 1}$, $\mathbf{h}_{1i}\in  \mathbb{C}^{N \times 1}$, $\mathbf{h}_{2r}\in  \mathbb{C}^{M \times 1}$, $\mathbf{h}_{2i}\in  \mathbb{C}^{N \times 1}$ and $\mathbf{H}_{ir}\in  \mathbb{C}^{M \times N}$ denote the frequency response of channels spanning from $\text{S}_1$ to AF relay, $\text{S}_1$ to IRS, $\text{S}_2$ to AF relay, $\text{S}_2$ to IRS and IRS to AF relay, respectively. $\bm{\Theta}_1 = \text{diag}(\alpha_{11} e^{j\theta_{11}}, \alpha_{12} e^{j\theta_{12}},\ldots,\alpha_{1N} e^{j\theta_{1N}})$ is the reflecting and configurable IRS matrix in the first time slot, where ${\alpha}_{1n} \in(0,1]$ and ${\theta}_{1n} \in (0,2\pi]$ respectively stand for the amplitude value and phase shift value of the $n$th reflection element. For simplicity, the amplitude value ${\alpha}_{1n}$ is generally set as 1. $\mathbf{n}_r$ denotes the received additive white Gaussian noise (AWGN) at AF relay with $\mathbf{n}_r \thicksim \mathcal{CN} (0,\sigma_r^2 \mathbf{I}_M)$.

In the second time slot, performing receive and transmit beamforming operations on the received signal. Accordingly, the processed signal can be written as
\begin{equation}\label{xr}
\mathbf{x}_r = \mathbf{A}\mathbf{y}_r,
\end{equation}
where $\mathbf{A} \in \mathbb{C}^{M \times M}$ is the beamforming matrix. AF relay transmit the processed signal to $\text{S}_1$ and $\text{S}_2$, while the transmit power of AF relay is given by
\begin{align}\label{pr}
P_r^t =  &P_1\| \mathbf{A}(\mathbf{h}_{1r}+\mathbf{H}_{ir}\bm{\Theta}_1\mathbf{h}_{1i})\|^2  \\
& + P_2\| \mathbf{A} (\mathbf{h}_{2r}+\mathbf{H}_{ir}\bm{\Theta}_1\mathbf{h}_{2i})\|^2 + \sigma_r^2 \|\mathbf{A}\|^2_F \nonumber\\
\leq & P_r, \nonumber
\end{align}
where $P_r$ represents the maximum transmit power of AF relay.
The received signal at \textcolor{blue}{$\text{S}_j$} ($j$=1, 2) can be written as
\begin{equation}\label{y01}
\textcolor{blue}{y_{0j} = (\mathbf{h}_{jr}^H+\mathbf{h}_{ji}^H \bm{\Theta}_2\mathbf{H}_{ir}^H)\mathbf{A}\mathbf{y}_r + n_j,}
%&= \sqrt{ P_1 }(\mathbf{h}_{1r}^H+\mathbf{h}_{1i}^H \bm{\Theta}_2\mathbf{H}_{ir}^H)\mathbf{A}(\mathbf{h}_{1r}+\mathbf{H}_{ir}\bm{\Theta}_1\mathbf{h}_{1i})x_1 \nonumber \\
%& + \sqrt{ P_2 }(\mathbf{h}_{1r}^H+\mathbf{h}_{1i}^H \bm{\Theta}_2\mathbf{H}_{ir}^H)\mathbf{A}(\mathbf{h}_{2r}+\mathbf{H}_{ir}\bm{\Theta}_1\mathbf{h}_{2i})x_2 \nonumber \\
%& +(\mathbf{h}_{1r}^H+\mathbf{h}_{1i}^H \bm{\Theta}_2\mathbf{H}_{ir}^H)\mathbf{A}\mathbf{n}_r+n_1,
\end{equation}
where $\bm{\Theta}_2 = \text{diag}(\alpha_{21} e^{j\theta_{21}}, \alpha_{22} e^{j\theta_{22}},\ldots,\alpha_{2N} e^{j\theta_{2N}})$ is the reflecting and configurable IRS matrix in the second time slot, where ${\alpha}_{2n} \in(0,1]$ and ${\theta}_{2n} \in (0,2\pi]$ are the amplitude value and phase shift value of the $n$th reflection element, respectively. Similarly, the amplitude value ${\alpha}_n$ is also equal to 1. \textcolor{blue}{$n_j$} is the received AWGN at \textcolor{blue}{$\text{S}_j$} with $\textcolor{blue}{n_j} \thicksim \mathcal{CN} (0, \textcolor{blue}{\sigma_j^2})$.
Since \textcolor{blue}{$\text{S}_j$} has perfect knowledge of the signal transmitted by itself, the self-interference can be eliminated by subtracting the term about \textcolor{blue}{$x_j$}. After the self-interference elimination, the equivalent received signal at \textcolor{blue}{$\text{S}_j$} can be obtained as follows
\begin{align}\label{y1}
\textcolor{blue}{	y_j = }& \textcolor{blue}{\sqrt{P_k}(\mathbf{h}_{jr}^H+\mathbf{h}_{ji}^H \bm{\Theta}_2\mathbf{H}_{ir}^H)\mathbf{A}(\mathbf{h}_{kr}+\mathbf{H}_{ir}\bm{\Theta}_1\mathbf{h}_{ki})x_k} \nonumber \\
& \textcolor{blue}{	+(\mathbf{h}_{jr}^H+\mathbf{h}_{ji}^H \bm{\Theta}_2\mathbf{H}_{ir}^H)\mathbf{A}\mathbf{n}_r + n_j,}
\end{align}
%\begin{align}\label{y1}
%y_1 = & \sqrt{P_2}(\mathbf{h}_{1r}^H+\mathbf{h}_{1i}^H \bm{\Theta}_2\mathbf{H}_{ir}^H)\mathbf{A}(\mathbf{h}_{2r}+\mathbf{H}_{ir}\bm{\Theta}_1\mathbf{h}_{2i})x_2 \nonumber \\
%& +(\mathbf{h}_{1r}^H+\mathbf{h}_{1i}^H \bm{\Theta}_2\mathbf{H}_{ir}^H)\mathbf{A}\mathbf{n}_r + n_1.
%\end{align}
%Similarly, the equivalent received signal at $\text{S}_2$ can be represented as follows
%\begin{align}\label{y2}
%y_2 = &\sqrt{P_1}(\mathbf{h}_{2r}^H+\mathbf{h}_{2i}^H \bm{\Theta}_2\mathbf{H}_{ir}^H)\mathbf{A}(\mathbf{h}_{1r}+\mathbf{H}_{ir}\bm{\Theta}_1\mathbf{h}_{1i})x_1 \nonumber \\
%&  +(\mathbf{h}_{2r}^H+\mathbf{h}_{2i}^H \bm{\Theta}_2\mathbf{H}_{ir}^H)\mathbf{A}\mathbf{n}_r + n_2,
%\end{align}
%where $n_2$ is the received AWGN at $\text{S}_2$ with $n_2 \thicksim \mathcal{CN} (0,\sigma_2^2)$.
\textcolor{blue}{where $(j, k)=\{(1, 2),(2, 1)\}$}.
The achievable rates at $\text{S}_2$ and $\text{S}_1$ can be respectively denoted as
\begin{align}
  R_{12}=\frac{1}{2}\log_2(1+\text{SNR}_{12}), \\
  R_{21}=\frac{1}{2}\log_2(1+\text{SNR}_{21}),
\end{align}
where $\text{SNR}_{12}$ and $\text{SNR}_{21}$ are the received $\text{SNRs}$ at $\text{S}_2$ (i.e., from $\text{S}_1$ to $\text{S}_2$ link) and $\text{S}_1$ (i.e., from $\text{S}_2$ to $\text{S}_1$ link). $\text{SNR}_{12}$ and $\text{SNR}_{21}$ can be respectively expressed as follow
\begin{align}
  &\text{SNR}_{12} = \nonumber\\
  &~\frac{ P_1 |(\mathbf{h}_{2r}^H+\mathbf{h}_{2i}^H \bm{\Theta}_2\mathbf{H}_{ir}^H)\mathbf{A}(\mathbf{h}_{1r}+\mathbf{H}_{ir}\bm{\Theta}_1\mathbf{h}_{1i})|^2}{ \|(\mathbf{h}_{2r}^H+\mathbf{h}_{2i}^H \bm{\Theta}_2\mathbf{H}_{ir}^H)\mathbf{A}\|^2 \sigma_r^2+ \sigma_2^2},  \\
  &\text{SNR}_{21} = \nonumber\\
  &~\frac{ P_2 |(\mathbf{h}_{1r}^H+\mathbf{h}_{1i}^H \bm{\Theta}_2\mathbf{H}_{ir}^H)\mathbf{A}(\mathbf{h}_{2r}+\mathbf{H}_{ir}\bm{\Theta}_1\mathbf{h}_{2i})|^2}{ \|(\mathbf{h}_{1r}^H+\mathbf{h}_{1i}^H \bm{\Theta}_2\mathbf{H}_{ir}^H)\mathbf{A}\|^2 \sigma_r^2+ \sigma_1^2}.
\end{align}
The system rate is defined as
\begin{equation}
R = \min\{R_{12},R_{21}\}.
\end{equation}

\subsection{Problem Formulation}
The optimization problem of maximizing system rate is cast as
\begin{subequations}\label{op1}
\begin{align}
& \max_{ \mathbf{A}, \bm{\Theta}_1, \bm{\Theta}_2 }~ \min\{R_{12},R_{21}\}  \\
& ~~~\text{s.t.} |\bm{\Theta}_1(n, n)|=1, |\bm{\Theta}_2(n, n)| =1, \forall n=1,\cdots,N,  \\
&~~~~~~~ P_1\| \mathbf{A}(\mathbf{h}_{1r}+\mathbf{H}_{ir}\bm{\Theta}_1\mathbf{h}_{1i})\|^2  \\
&~~~~~~ + P_2\| \mathbf{A} (\mathbf{h}_{2r}+\mathbf{H}_{ir}\bm{\Theta}_1\mathbf{h}_{2i})\|^2 + \sigma_r^2 \|\mathbf{A}\|^2_F \leq P_r. \label{P_r} \nonumber
\end{align}
\end{subequations}
\textcolor{blue}{In order to handle the above problem more conveniently}, defining $\bm{\theta}_1 \textcolor{blue}{\triangleq} [e^{j\theta_{11}},e^{j\theta_{12}},\ldots,e^{j\theta_{1N}}]^T$, $\bar{\bm{\theta}}_1=[\bm{\theta}_1; 1]$, $\bm{\theta}_2 \textcolor{blue}{\triangleq} [e^{j\theta_{21}},e^{j\theta_{22}},\ldots,e^{j\theta_{2N}}]^H$ and $\bar{\bm{\theta}}_2=[\bm{\theta}_2;1]$. We have
\begin{subequations}
\begin{align}
&\textcolor{blue}{\mathbf{h}_{jr} + \mathbf{H}_{ir}\bm{\Theta}_1\mathbf{h}_{ji} = \mathbf{H}_j\bar{\bm{\theta}}_1,} \\
%&\textcolor{blue}{\mathbf{h}_{2r} + \mathbf{H}_{ir}\bm{\Theta}_1\mathbf{h}_{2i} = \mathbf{H}_2\bar{\bm{\theta}}_1,} \\
&\textcolor{blue}{\mathbf{h}_{jr}^H + \mathbf{h}_{ji}^H \bm{\Theta}_2 \mathbf{H}_{ir}^H = \bar{\bm{\theta}}_2^H \mathbf{H}_j^H,}
%&\textcolor{blue}{\mathbf{h}_{2r}^H + \mathbf{h}_{2i}^H \bm{\Theta}_2 \mathbf{H}_{ir}^H = \bar{\bm{\theta}}_2^H \mathbf{H}_2^H,}
\end{align}
\end{subequations}
%\begin{subequations}
%	\begin{align}
%		&\textcolor{blue}{\mathbf{h}_{1r} + \mathbf{H}_{ir}\bm{\Theta}_1\mathbf{h}_{1i} = \mathbf{H}_1\bar{\bm{\theta}}_1,} \\
%		&\textcolor{blue}{\mathbf{h}_{2r} + \mathbf{H}_{ir}\bm{\Theta}_1\mathbf{h}_{2i} = \mathbf{H}_2\bar{\bm{\theta}}_1,} \\
%		&\textcolor{blue}{\mathbf{h}_{1r}^H + \mathbf{h}_{1i}^H \bm{\Theta}_2 \mathbf{H}_{ir}^H = \bar{\bm{\theta}}_2^H \mathbf{H}_1^H,} \\
%		&\textcolor{blue}{\mathbf{h}_{2r}^H + \mathbf{h}_{2i}^H \bm{\Theta}_2 \mathbf{H}_{ir}^H = \bar{\bm{\theta}}_2^H \mathbf{H}_2^H,}
%	\end{align}
%\end{subequations}
\textcolor{blue}{where $\mathbf{H}_j = [\mathbf{H}_{ir}\text{diag}(\mathbf{h}_{ji}),\mathbf{h}_{jr}]$}.
% and
%$\mathbf{H}_2 = [\mathbf{H}_{ir}\text{diag}(\mathbf{h}_{2i}),\mathbf{h}_{2r}]$.
Accordingly, the optimization problem can be reformulated as
\begin{subequations}\label{op2}
\begin{align}
&\max_{\mathbf{A},\bar{\bm{\theta}}_1,\bar{\bm{\theta}}_2}~ \min\{R_{12},R_{21}\} \\
&~  \text{s.t.}   |\bar{\bm{\theta}}_1(n)|=1, |\bar{\bm{\theta}}_2(n)| =1, \forall n=1,\cdots,N, \label{op2_2}\\
&~~~~~  \bar{\bm{\theta}}_1(N+1)=1,~ \bar{\bm{\theta}}_2(N+1) =1, \label{op2_2_1}\\
&~~~~  P_1\| \mathbf{A}\mathbf{H}_1\bar{\bm{\theta}}_1\|^2 + P_2\| \mathbf{A}\mathbf{H}_2\bar{\bm{\theta}}_1\|^2  + \sigma_r^2 \|\mathbf{A}\|_F^2\leq P_r, \label{op2_3}
\end{align}
\end{subequations}
where
\begin{align}
&R_{12} = \frac{1}{2}\log_2(1+\frac{ P_1 |\bar{\bm{\theta}}_2^H \mathbf{H}_2^H \mathbf{A} \mathbf{H}_1\bar{\bm{\theta}}_1|^2}{ \|\bar{\bm{\theta}}_2^H \mathbf{H}_2^H \mathbf{A}\|^2 \sigma_r^2+ \sigma_2^2}), \label{rr1}\\
&R_{21} = \frac{1}{2}\log_2(1+\frac{ P_2 |\bar{\bm{\theta}}_2^H \mathbf{H}_1^H \mathbf{A} \mathbf{H}_2\bar{\bm{\theta}}_1|^2}{ \|\bar{\bm{\theta}}_2^H \mathbf{H}_1^H \mathbf{A}\|^2 \sigma_r^2+ \sigma_1^2}). \label{rr2}
\end{align}

Let us define a variable $t$, $R_{12} \geq t$ and $R_{21} \geq t$. According to the property of logarithmic function, problem (\ref{op2}) can be converted to
\begin{subequations}\label{op3}
\begin{align}
&\max_{t,\mathbf{A},\bar{\bm{\theta}}_1,\bar{\bm{\theta}}_2}~~ t \\
&~~~~  \text{s.t.}~~~  2^{2t} \leq 1+ \frac{P_1|\bar{\bm{\theta}}_2^H \mathbf{H}_2^H \mathbf{A} \mathbf{H}_1\bar{\bm{\theta}}_1|^2}{\|\bar{\bm{\theta}}_2^H \mathbf{H}_2^H \mathbf{A}\|^2 \sigma_r^2+ \sigma_2^2}, \label{op3_1}\\
&~~~~~~~~~~~~  2^{2t} \leq 1+ \frac{P_2|\bar{\bm{\theta}}_2^H \mathbf{H}_1^H \mathbf{A} \mathbf{H}_2\bar{\bm{\theta}}_1|^2}{\|\bar{\bm{\theta}}_2^H \mathbf{H}_1^H \mathbf{A}\|^2 \sigma_r^2+ \sigma_1^2}, \label{op3_2}\\
&~~~~~~~~~~~~  \text{(\ref{op2_2})},~ \text{(\ref{op2_2_1})},~ \text{(\ref{op2_3})}.
\end{align}
\end{subequations}
It is too difficult to be solved directly because the variables, i.e., the beamforming matrix $\mathbf{A}$, the phase shift vectors $\bar{\bm{\theta}}_1$ and $\bar{\bm{\theta}}_2$, are coupled with each other, which make the problem more intractable. Here, there exist two alternating iteration schemes, namely LC-ZF-SCA and ONS-SDP-PSCA are proposed to tackle the optimization problem, and the related details are as follow.

\section{Proposed LC-ZF-SCA-based Method} \label{LC-ZF-SCA}
In the section, optimizing a variable or a group of variables given the remaining ones, there exist three subproblems need to be dealt with. In this case, a LC-ZF-SCA-based scheme is proposed to alternately optimize the variables $\mathbf{A}$, $\bar{\bm{\theta}}_1$ and $\bar{\bm{\theta}}_2$ for maximum rate. Here, ZF scheme is utilised to address the subproblem of calculating $\mathbf{A}$, where $\mathbf{A}$ can be obtained in closed form. For $\bar{\bm{\theta}}_1$ and $\bar{\bm{\theta}}_2$, we apply SCA algorithm to solve the corresponding two subproblems.

\subsection{Optimization of \ \bf{A} \ }
This subsection is aiming at obtaining beamforming matrix $\mathbf{A}$ at AF relay with given $\bar{\bm{\theta}}_1$ and $\bar{\bm{\theta}}_2$. Since the computational complexity of directly optimizing the beamforming matrix is extremely high, we adopt a low-complexity ZF scheme to obtain $\mathbf{A}$. Defining $\bar{\mathbf{H}}_1 \triangleq [\mathbf{H}_1\bar{\bm{\theta}}_1,\mathbf{H}_2\bar{\bm{\theta}}_1]$ and $\bar{\mathbf{H}}_2 \triangleq [\mathbf{H}_2\bar{\bm{\theta}}_2,\mathbf{H}_1\bar{\bm{\theta}}_2]^H$, in the light of ZF criterion \cite{Lk2014}, $\mathbf{A}$ can be denoted as
\begin{equation}\label{A}
\mathbf{A} = \tau\bar{\mathbf{H}}^{\dag}_2\bar{\mathbf{H}}^{\dag}_1.
\end{equation}
Plugging (\ref{A}) into the transmit power constraint with equality, we have $\tau$
as (\ref{tau}), as shown at the top of next page.
 \begin{figure*}[ht] %hb´ú±í·ÅÔÚÎÄÕµײ¿£¬%htΪ·ÅÔÚÎÄÕ¶¥²¿
	\centering
	\begin{equation}\label{tau}
		\tau = \sqrt{\frac{P_r}{ P_1\| \bar{\mathbf{H}}^{\dag}_2\bar{\mathbf{H}}^{\dag}_1\mathbf{H}_1 \bar{\bm{\theta}}_1\|^2+P_2 \|\bar{\mathbf{H}}^{\dag}_2\bar{\mathbf{H}}^{\dag}_1 \mathbf{H}_2 \bar{\bm{\theta}}_1\|^2+\sigma_r^2 \|\bar{\mathbf{H}}^{\dag}_2\bar{\mathbf{H}}^{\dag}_1\|^2_F}}.
	\end{equation}
	\hrulefill
\end{figure*}
%\begin{align}\label{tau}
%&\tau = \\
%&\sqrt{\frac{P_r}{ P_1\| \bar{\mathbf{H}}^{\dag}_2\bar{\mathbf{H}}^{\dag}_1\mathbf{H}_1 \bar{\bm{\theta}}_1\|^2+P_2 \|\bar{\mathbf{H}}^{\dag}_2\bar{\mathbf{H}}^{\dag}_1 \mathbf{H}_2 \bar{\bm{\theta}}_1\|^2+\sigma_r^2 \|\bar{\mathbf{H}}^{\dag}_2\bar{\mathbf{H}}^{\dag}_1\|^2_F}}. \nonumber
%\end{align}
Thereby, the AF beamforming matrix $\mathbf{A}$ can be obtained.

\subsection{Optimization of $\bar{\bm{\theta}}_1$}
%\subsection{Optimization of  $\overline{\bm{\theta}}_1$ }

Fixing $\bar{\bm{\theta}}_2$ and $\mathbf{A}$, problem (\ref{op3}) can be transformed to
\begin{subequations}\label{op5}
\begin{align}
&\max_{t, \bar{\bm{\theta}}_1} ~~t  \\
&~\text{s.t.}~~~ |\bar{\bm{\theta}}_1(n)|=1,~ \forall n=1,\cdots,N, \label{op5_1} \\
&~~~~~~~~~ \bar{\bm{\theta}}_1(N+1)=1,  \label{op5_5}\\
&~~~~~~~~~ 2^{2t} \leq 1+ \frac{\bar{\bm{\theta}}_1^H \mathbf{B}_{12} \bar{\bm{\theta}}_1 }{\|\bar{\bm{\theta}}_2^H \mathbf{H}_2^H \mathbf{A}\|^2 \sigma_r^2+ \sigma_2^2}, \label{op5_2} \\
&~~~~~~~~~ 2^{2t} \leq 1+ \frac{\bar{\bm{\theta}}_1^H \mathbf{B}_{21} \bar{\bm{\theta}}_1 }{\|\bar{\bm{\theta}}_2^H \mathbf{H}_1^H \mathbf{A}\|^2 \sigma_r^2+ \sigma_1^2}, \label{op5_3} \\
&~~~~~~~~~ \bar{\bm{\theta}}_1^H(P_1\mathbf{H}_1^H \mathbf{A}^H \mathbf{A}\mathbf{H}_1 + P_2\mathbf{H}_2^H \mathbf{A}^H \mathbf{A}\mathbf{H}_2)\bar{\bm{\theta}}_1 \nonumber \\
&~~~~~~~~~ +\sigma_r^2 \|\mathbf{A}\|^2_F\leq P_r, \label{op5_4}
\end{align}
\end{subequations}
where matrices $\mathbf{B}_{12}=P_1 \mathbf{H}_1^H \mathbf{A}^H \mathbf{H}_2 \bar{\bm{\theta}}_2 \bar{\bm{\theta}}_2^H \mathbf{H}_2^H \mathbf{A} \mathbf{H}_1$ and
$\mathbf{B}_{21}=P_2 \mathbf{H}_2^H \mathbf{A}^H \mathbf{H}_1 \bar{\bm{\theta}}_2 \bar{\bm{\theta}}_2^H \mathbf{H}_1^H \mathbf{A} \mathbf{H}_2$. Constraints (\ref{op5_1}), (\ref{op5_2}) and (\ref{op5_3}) are non-convex.
Relaxing constraint (\ref{op5_1}), we can obtain
\begin{equation}\label{op5_1_1}
\bar{\bm{\theta}}_1^H(n) \bar{\bm{\theta}}_1(n) \leq 1,~ \forall n=1,\cdots,N,
\end{equation}
which is a convex constraint.
In order to convert the constraint (\ref{op5_2}) from non-convex to convex, the first-order Taylor approximation can be used to obtain the low bound of the numerator of fraction in (\ref{op5_2}). Its corresponding first-order Taylor expansion can be performed as follows
\begin{equation}
\bar{\bm{\theta}}_1^H \mathbf{B}_{12} \bar{\bm{\theta}}_1 \geq 2 \mathfrak{R}\{\tilde{\bm{\theta}}_1^H \mathbf{B}_{12} \bar{\bm{\theta}}_1\} -  \tilde{\bm{\theta}}_1^H \mathbf{B}_{12} \tilde{\bm{\theta}}_1,  \label{op5_2_1}
\end{equation}
where $\tilde{\bm{\theta}}_1$ is the feasible point. Similarly, performing first-order Taylor approximation operation on the numerator of fraction in (\ref{op5_3}), we have
\begin{equation}
\bar{\bm{\theta}}_1^H \mathbf{B}_{21} \bar{\bm{\theta}}_1 \geq 2 \mathfrak{R}\{\tilde{\bm{\theta}}_1^H \mathbf{B}_{21} \bar{\bm{\theta}}_1\} - \tilde{\bm{\theta}}_1^H \mathbf{B}_{21} \tilde{\bm{\theta}}_1. \label{op5_3_1}
\end{equation}
Substituting (\ref{op5_2_1}) and (\ref{op5_3_1}) into constraint (\ref{op5_2}) and (\ref{op5_3}), the constraints can be respectively rewritten as
\begin{align}
  2^{2t} &\leq 1+ \frac{2 \mathfrak{R}\{\tilde{\bm{\theta}}_1^H \mathbf{B}_{12} \bar{\bm{\theta}}_1\} -  \tilde{\bm{\theta}}_1^H \mathbf{B}_{12} \tilde{\bm{\theta}}_1}{\|\bar{\bm{\theta}}_2^H \mathbf{H}_2^H \mathbf{A}\|^2 \sigma_r^2+ \sigma_2^2}, \label{op5_2_1_1} \\
  2^{2t} &\leq 1+ \frac{2 \mathfrak{R}\{\tilde{\bm{\theta}}_1^H \mathbf{B}_{21} \bar{\bm{\theta}}_1\} -  \tilde{\bm{\theta}}_1^H \mathbf{B}_{21} \tilde{\bm{\theta}}_1}{\|\bar{\bm{\theta}}_2^H \mathbf{H}_1^H \mathbf{A}\|^2 \sigma_r^2+ \sigma_1^2}, \label{op5_3_1_1}
\end{align}
which are convex constraints.
Therefore, the optimization problem (\ref{op5}) can be transformed to
\begin{subequations}\label{op6}
\begin{align}
&\max_{t, \bar{\bm{\theta}}_1}~~~~  t \\
&~~ \text{s.t.}~~~~ \text{(\ref{op5_5})},~\text{(\ref{op5_4})},~ \text{(\ref{op5_1_1})},~ \text{(\ref{op5_2_1_1})},~ \text{(\ref{op5_3_1_1})}.
\end{align}
\end{subequations}
Because of a linear objective function and several convex constraints, the above problem is convex, which can be solved by optimization solver, such as CVX.
So the solution $\bm{\theta}_1$ is calculated as
\begin{equation}
\bm{\theta}_1=e^{j\text{arg}[{\frac{\bar{\bm{\theta}}_1}{\bar{\bm{\theta}}_1(N+1)}(1:N)}]}.
\end{equation}

\subsection{ Optimization of $\bar{\bm{\theta}}_2$ }
When $\bar{\bm{\theta}}_1$ and $\mathbf{A}$ are given, the optimization problem (\ref{op3}) can be reduced to
\begin{subequations}\label{op7}
\begin{align}
&\max_{t,\bar{\bm{\theta}}_2}~~  t  \\
&~\text{s.t.}~~~|\bar{\bm{\theta}}_2(n)|=1,~ \forall n=1,\cdots,N, \label{op7_1}\\
&~~~~~~~~~ \bar{\bm{\theta}}_2(N+1)=1,  \label{op7_4}\\
&~~~~~~~~~ 2^{2t} \leq 1+ \frac{P_1|\bar{\bm{\theta}}_2^H \mathbf{H}_2^H \mathbf{A} \mathbf{H}_1\bar{\bm{\theta}}_1|^2}{\|\bar{\bm{\theta}}_2^H \mathbf{H}_2^H \mathbf{A}\|^2 \sigma_r^2+ \sigma_2^2}, \label{op7_2} \\
&~~~~~~~~~ 2^{2t} \leq 1+ \frac{P_2|\bar{\bm{\theta}}_2^H \mathbf{H}_1^H \mathbf{A} \mathbf{H}_2\bar{\bm{\theta}}_1|^2}{\|\bar{\bm{\theta}}_2^H \mathbf{H}_1^H \mathbf{A}\|^2 \sigma_r^2+ \sigma_1^2}, \label{op7_3}
\end{align}
\end{subequations}
which is non-convex because of the non-convex constraints (\ref{op7_1}), (\ref{op7_2}) and (\ref{op7_3}).
For constraint (\ref{op7_1}), a relaxation strategy similar to (\ref{op5_1_1}) can be performed as follows
\begin{equation}\label{op7_1_1}
\bar{\bm{\theta}}_2^H(n) \bar{\bm{\theta}}_2(n) \leq 1 , ~ \forall n=1,\cdots,N,
\end{equation}
which is a convex constraint.
For constraint (\ref{op7_2}), its fraction part can be rewritten as follows
\begin{align}\label{op7_2_1}
\frac{P_1|\bar{\bm{\theta}}_2^H \mathbf{H}_2^H \mathbf{A} \mathbf{H}_1\bar{\bm{\theta}}_1|^2}{\|\bar{\bm{\theta}}_2^H \mathbf{H}_2^H \mathbf{A}\|^2 \sigma_r^2+ \sigma_2^2}
%&~~~~~~~~~~~~~~ =\frac{\beta_1 P \bar{\bm{\theta}}_2^H \mathbf{H}_2^H \mathbf{A} \mathbf{H}_1\bar{\bm{\theta}}_1 \bar{\bm{\theta}}_1^H  \mathbf{H}_1^H \mathbf{A}^H \mathbf{H}_2 \bar{\bm{\theta}}_2 }{\sigma_r^2\bar{\bm{\theta}}_2^H\mathbf{H}_2^H \mathbf{A}\mathbf{A}^H \mathbf{H}_2 \bar{\bm{\theta}}_2 + \sigma_2^2} \nonumber \\
=\frac{\bar{\bm{\theta}}_2^H\mathbf{C}_{12}\bar{\bm{\theta}}_2}{\bar{\bm{\theta}}_2^H\mathbf{D}_{12}\bar{\bm{\theta}}_2},
\end{align}
where matrices $\mathbf{C}_{12}=P_1  \mathbf{H}_2^H \mathbf{A} \mathbf{H}_1\bar{\bm{\theta}}_1 \bar{\bm{\theta}}_1^H  \mathbf{H}_1^H \mathbf{A}^H \mathbf{H}_2$ and $\mathbf{D}_{12}=\sigma_r^2\mathbf{H}^H_2 \mathbf{A}\mathbf{A}^H \mathbf{H}_2 +
\left[
\begin{array}{cc}
\mathbf{0}_{N \times N} & \mathbf{0}_{N \times 1} \\
\mathbf{0}_{1 \times N} & \sigma_2^2 \\
    \end{array}
  \right]$.
Substituting (\ref{op7_2_1}) into (\ref{op7_2}) yields the following inequality
\begin{equation}\label{op7_2_2}
2^{2t} \leq 1+ \frac{\bar{\bm{\theta}}_2^H\mathbf{C}_{12}\bar{\bm{\theta}}_2}{\bar{\bm{\theta}}_2^H\mathbf{D}_{12}\bar{\bm{\theta}}_2},
\end{equation}
it is found that the above constraint is still non-convex.
According to \cite{Gxr2020}, the first-order Taylor expansion of (\ref{op7_2_1}) at a feasible point $\tilde{\bm{\theta}}_2$ can be expressed as follows
\begin{equation}\label{op7_2_3}
\frac{\bar{\bm{\theta}}_2^H\mathbf{C}_{12}\bar{\bm{\theta}}_2}{\bar{\bm{\theta}}_2^H\mathbf{D}_{12}\bar{\bm{\theta}}_2} \geq 2\mathfrak{R}\{\mathbf{f}_{12}^H \bar{\bm{\theta}}_2\} + d_{12},
\end{equation}
where vector
$\mathbf{f}_{12}^H =
\frac{\tilde{\bm{\theta}}_2^H \mathbf{C}_{12}}{\tilde{\bm{\theta}}_2^H \mathbf{D}_{12} \tilde{\bm{\theta}}_2}-\tilde{\bm{\theta}}_2^H(\mathbf{D}_{12}-\lambda_{max}^{\mathbf{D}_{12}}\mathbf{I}_{N+1})\frac{\tilde{\bm{\theta}}_2^H \mathbf{C}_{12} \tilde{\bm{\theta}}_2}{(\tilde{\bm{\theta}}_2^H \mathbf{D}_{12} \tilde{\bm{\theta}}_2)^2}$,
$d_{12} = -[2\lambda_{max}^{\mathbf{D}_{12}}(N+1) - \tilde{\bm{\theta}}_2^H \mathbf{D}_{12} \tilde{\bm{\theta}}_2] \frac{\tilde{\bm{\theta}}_2^H \mathbf{C}_{12} \tilde{\bm{\theta}}_2}{(\tilde{\bm{\theta}}_2^H \mathbf{D}_{12} \tilde{\bm{\theta}}_2)^2}$
%\begin{align}
%&\mathbf{f}_{12}^H =  \label{f12}\\
%&\frac{\tilde{\bm{\theta}}_2^H \mathbf{C}_{12}}{\tilde{\bm{\theta}}_2^H \mathbf{D}_{12} \tilde{\bm{\theta}}_2}-\tilde{\bm{\theta}}_2^H(\mathbf{D}_{12}-\lambda_{max}^{\mathbf{D}_{12}}\mathbf{I}_{N+1})\frac{\tilde{\bm{\theta}}_2^H \mathbf{C}_{12} \tilde{\bm{\theta}}_2}{(\tilde{\bm{\theta}}_2^H \mathbf{D}_{12} \tilde{\bm{\theta}}_2)^2}, \nonumber  \\
%&d_{12} = -[2\lambda_{max}^{\mathbf{D}_{12}}(N+1) - \tilde{\bm{\theta}}_2^H \mathbf{D}_{12} \tilde{\bm{\theta}}_2] \frac{\tilde{\bm{\theta}}_2^H \mathbf{C}_{12} \tilde{\bm{\theta}}_2}{(\tilde{\bm{\theta}}_2^H \mathbf{D}_{12} \tilde{\bm{\theta}}_2)^2}, \label{d12}
%\end{align}
and $\lambda_{max}^{\mathbf{D}_{12}}$ is the maximum eigenvalue of $\mathbf{D}_{12}$.
Substituting the low bound of $\frac{\bar{\bm{\theta}}_2^H\mathbf{C}_{12}\bar{\bm{\theta}}_2}{\bar{\bm{\theta}}_2^H\mathbf{D}_{12}\bar{\bm{\theta}}_2}$ into (\ref{op7_2_2}) yields
\begin{equation}\label{op7_2_4}
2^{2t} \leq 1+ 2\mathfrak{R}\{\mathbf{f}_{12}^H \bar{\bm{\theta}}_2\} + d_{12},
\end{equation}
which is convex.
In the same manner, for the fraction part of constraint (\ref{op7_3}) we have
\begin{align}\label{op7_3_1}
\frac{P_2|\bar{\bm{\theta}}_2^H \mathbf{H}_1^H \mathbf{A} \mathbf{H}_2\bar{\bm{\theta}}_1|^2}{\|\bar{\bm{\theta}}_2^H \mathbf{H}_1^H \mathbf{A}\|^2 \sigma_r^2+ \sigma_1^2} &=\frac{\bar{\bm{\theta}}_2^H\mathbf{C}_{21}\bar{\bm{\theta}}_2}{\bar{\bm{\theta}}_2^H\mathbf{D}_{21}\bar{\bm{\theta}}_2} \nonumber\\
& \geq 2\mathfrak{R}\{\mathbf{f}_{21}^H \bar{\bm{\theta}}_2\} + d_{21},
\end{align}
where matrices $\mathbf{C}_{21} = P_2  \mathbf{H}_1^H \mathbf{A} \mathbf{H}_2\bar{\bm{\theta}}_1 \bar{\bm{\theta}}_1^H  \mathbf{H}_2^H \mathbf{A}^H \mathbf{H}_1$, $\mathbf{D}_{21}=\sigma_r^2\mathbf{H}_1^H \mathbf{A}\mathbf{A}^H \mathbf{H}_1 +
\left[
\begin{array}{cc}
\mathbf{0}_{N \times N} & \mathbf{0}_{N \times 1} \\
\mathbf{0}_{1 \times N} & \sigma_1^2 \\
    \end{array}
  \right]$,
$\mathbf{f}_{21}^H = \frac{\tilde{\bm{\theta}}_2^H \mathbf{C}_{21}}{\tilde{\bm{\theta}}_2^H \mathbf{D}_{21} \tilde{\bm{\theta}}_2} -\tilde{\bm{\theta}}_2^H(\mathbf{D}_{21}-\lambda_{max}^{\mathbf{D}_{21}}\mathbf{I}_{N+1})\frac{\tilde{\bm{\theta}}_2^H \mathbf{C}_{21} \tilde{\bm{\theta}}_2}{(\tilde{\bm{\theta}}_2^H \mathbf{D}_{21} \tilde{\bm{\theta}}_2)^2}$ and $d_{21} = -[2\lambda_{max}^{\mathbf{D}_{21}}(N+1) - \tilde{\bm{\theta}}_2^H \mathbf{D}_{21} \tilde{\bm{\theta}}_2] \frac{\tilde{\bm{\theta}}_2^H \mathbf{C}_{21} \tilde{\bm{\theta}}_2}{(\tilde{\bm{\theta}}_2^H \mathbf{D}_{21} \tilde{\bm{\theta}}_2)^2}$.
%\begin{align}
%&\mathbf{f}_{21}^H =  \label{f21} \\
%&\frac{\tilde{\bm{\theta}}_2^H \mathbf{C}_{21}}{\tilde{\bm{\theta}}_2^H \mathbf{D}_{21} \tilde{\bm{\theta}}_2} -\tilde{\bm{\theta}}_2^H(\mathbf{D}_{21}-\lambda_{max}^{\mathbf{D}_{21}}\mathbf{I}_{N+1})\frac{\tilde{\bm{\theta}}_2^H \mathbf{C}_{21} \tilde{\bm{\theta}}_2}{(\tilde{\bm{\theta}}_2^H \mathbf{D}_{21} \tilde{\bm{\theta}}_2)^2},   \nonumber\\  %\label{f21}
%&d_{21} = -[2\lambda_{max}^{\mathbf{D}_{21}}(N+1) - \tilde{\bm{\theta}}_2^H \mathbf{D}_{21} \tilde{\bm{\theta}}_2] \frac{\tilde{\bm{\theta}}_2^H \mathbf{C}_{21} \tilde{\bm{\theta}}_2}{(\tilde{\bm{\theta}}_2^H \mathbf{D}_{21} \tilde{\bm{\theta}}_2)^2}. \label{d21}
%\end{align}
Correspondingly, constraint (\ref{op7_3}) can be transformed to be convex as follows
\begin{equation}\label{op7_3_2}
2^{2t} \leq 1+ 2\mathfrak{R}\{\mathbf{f}_{21}^H \bar{\bm{\theta}}_2\} + d_{21}.
\end{equation}
Therefore, the optimization problem (\ref{op7}) can be reformulated as follows
\begin{subequations}\label{op8}
\begin{align}
& \max_{t,\bar{\bm{\theta}}_2}~~~~  t  \\
&~ \text{s.t.}~~~~~ \text{(\ref{op7_4})},~ \text{(\ref{op7_1_1})},~ \text{(\ref{op7_2_4})},~ \text{(\ref{op7_3_2})}.
\end{align}
\end{subequations}
The solution $\bar{\bm{\theta}}_2$ can be directly solved by CVX optimization tool. Thereby, the solution $\bm{\theta}_2$ can be achieved as
\begin{equation}
\bm{\theta}_2=e^{j\text{arg}[{\frac{\bar{\bm{\theta}}_2}{\bar{\bm{\theta}}_2(N+1)}(1:N)}]}.
\end{equation}

\subsection{Overall algorithm}
The optimization problem have \textcolor{blue}{an} upper bound because of the non-decreasing property and limited transmit power of $\text{S}_1$, $\text{S}_2$ and AF relay.
Performing alternative iteration among $\mathbf{A}$, $\bar{\bm{\theta}}_1$ and $\bar{\bm{\theta}}_2$  until the convergence criterion is satisfied. The proposed LC-ZF-SCA algorithm is summarized in the following Algorithm 1.
\begin{algorithm}
	\caption{Proposed LC-ZF-SCA algorithm}
	\begin{algorithmic}[1]
		\STATE Initialize $\mathbf{A}^0$, $\bar{\bm{\theta}}_1^0$ and $\bar{\bm{\theta}}_2^0$, $R^0$ can be calculated.
		\STATE Set the convergence error $\delta$ and initialize iteration number $k=0$.
		\REPEAT
		\STATE Given ($\bar{\bm{\theta}}_1^{k}$, $\bar{\bm{\theta}}_2^{k}$), calculate $\mathbf{A}^{k+1}$ with (\ref{A}) and (\ref{tau}).
		\STATE Given ($\mathbf{A}^{k+1}$, $\bar{\bm{\theta}}_2^{k}$), solve problem (\ref{op6}) to obtain $\bar{\bm{\theta}}_1^{k+1}$.
        \STATE Given ($\mathbf{A}^{k+1}$, $\bar{\bm{\theta}}_1^{k+1}$), solve problem (\ref{op8}) to obtain $\bar{\bm{\theta}}_2^{k+1}$.
        \STATE Calculate $R^{k+1}$ by using ($\mathbf{A}^{k+1}$, $\bar{\bm{\theta}}_1^{k+1}$, $\bar{\bm{\theta}}_2^{k+1}$).
		\STATE $k=k+1$.
		\UNTIL
		\STATE $|R^{k+1} - R^k| \leq \delta$.
	\end{algorithmic}
\end{algorithm}

The computational complexity of LC-ZF-SCA algorithm contains three parts related to $\mathbf{A}$, $\bar{\bm{\theta}}_1$ and $\bar{\bm{\theta}}_2$. The complexity of $\mathbf{A}$ is denoted as $M^3+11M^2+10MN+7M+6$ float-point operations (FLOPs) in line with (\ref{A}) and (\ref{tau}). Since problem (\ref{op8}) has three linear constraints, one second-order cone (SOC) constraint of dimension $N+1$ and one SOC constraint of dimension $N$, the complexity of calculating $\bar{\bm{\theta}}_1$ is denoted as $n_{\bar{\bm{\theta}}_1}\sqrt{7}[(N+1)^2+N^2+n_{\bar{\bm{\theta}}_1}^2+3n_{\bar{\bm{\theta}}_1}+3]$, where $n_{\bar{\bm{\theta}}_1}=N+2$ is the number of variables. Besides, there are three linear constraints, one SOC constraint of dimension $N$ and $N+2$ variables in problem (\ref{op8}), it complexity is presented as $n_{\bar{\bm{\theta}}_2}\sqrt{5}[N^2+n_{\bar{\bm{\theta}}_2}^2+3n_{\bar{\bm{\theta}}_2}+3]$, where $n_{\bar{\bm{\theta}}_2}=N+2$. Therefore, the complexity of the proposed LC-ZF-SCA algorithm can be written as follows
\begin{align}
&\mathcal{O}\{D_1[M^3+11M^2+10MN+7M+6 \nonumber\\
&+n_{\bar{\bm{\theta}}_1}\sqrt{7}((N+1)^2+N^2+n_{\bar{\bm{\theta}}_1}^2+3n_{\bar{\bm{\theta}}_1}+3) \nonumber\\
&+n_{\bar{\bm{\theta}}_2}\sqrt{5}(N^2+n_{\bar{\bm{\theta}}_2}^2+3n_{\bar{\bm{\theta}}_2}+3)
]\text{ln}(1/\varepsilon)\}
\end{align}
FLOPs, where $D_1$ is the iterative number in Algorithm 1 and $\varepsilon$ is the computation accuracy.
%It is noted that the highest order of computational complexity is $M^3$ and $N^3$ FLOPs.

\section{Proposed ONS-SDP-PSCA-based Method} \label{ONS-SDP-PSCA}
In the section III, the LC-ZF-SCA-based scheme is put forward to optimize AF relay beamforming matrix $\mathbf{A}$, IRS reflecting coefficient vectors $\bar{\bm{\theta}}_1$ and $\bar{\bm{\theta}}_2$. To obtain a rate performance improvement, a high-performance ONS-SDP-PSCA-based scheme is proposed, where the subproblem related to $\mathbf{A}$ is firstly solved by SVD and ONS method. Then the optimization problem is translated into two SDP subproblems, where IRS phase shift matrix in the second time slot is optimized by GFP algorithm, and a penalty function is adopted to recover rank-one IRS phase shift matrices of two time slots. The associated details are presented as follow.

\subsection{ Optimization of \ \bf{A} \ }\label{section_IV_A}

When IRS reflecting coefficient vectors $\bar{\bm{\theta}}_1$ and $\bar{\bm{\theta}}_2$ are fixed, the SVD of $\bar{\mathbf{H}}_1$ and $\bar{\mathbf{H}}_2$ can be respectively expressed as follow
\begin{align}
\bar{\mathbf{H}}_1 &= \mathbf{U}_1\bm{\Sigma}_1\mathbf{V}_1^H \\
&= [\mathbf{U}_{11}~ \mathbf{U}_{12}]\left[
\begin{array}{cc}
\bm{\Sigma}_{11} \\
\bm{\Sigma}_{12} \\
\end{array}
\right]\mathbf{V}_1^H \nonumber\\
&= \mathbf{U}_{11}\bm{\Sigma}_{11}\mathbf{V}_1^H, \nonumber\\
\bar{\mathbf{H}}_2 &= \mathbf{U}_2\bm{\Sigma}_2\mathbf{V}_2^H \\
&= \mathbf{U}_2[\bm{\Sigma}_{21}~ \bm{\Sigma}_{22}]\left[
\begin{array}{cc}
\mathbf{V}_{21}^H \\
\mathbf{V}_{22}^H \\
\end{array}
\right] \nonumber\\
&= \mathbf{U}_2\bm{\Sigma}_{21}\mathbf{V}_{21}^H, \nonumber
\end{align}
where $\mathbf{U}_1 \in \mathbb{C}^{M \times M}$, $\mathbf{V}_1 \in \mathbb{C}^{2 \times 2}$, $\mathbf{U}_2 \in \mathbb{C}^{2 \times 2}$, and $\mathbf{V}_2 \in \mathbb{C}^{M \times M}$ are the unitary matrices, $\bm{\Sigma}_1 \in \mathbb{C}^{M \times 2}$ and $\bm{\Sigma}_2 \in \mathbb{C}^{2 \times M}$ are the matrices with singular values for elements on the main diagonal and 0 for other elements. $\mathbf{U}_{11} \in \mathbb{C}^{M \times 2}$, $\mathbf{U}_{12} \in \mathbb{C}^{M \times (M-2)}$, $\mathbf{V}_{21} \in \mathbb{C}^{M \times 2}$ and $\mathbf{V}_{22} \in \mathbb{C}^{M \times (M-2)}$, $\bm{\Sigma}_{11} \in \mathbb{C}^{2 \times 2}$ and $\bm{\Sigma}_{21} \in \mathbb{C}^{2 \times 2}$ are diagonal matrices consisting of singular values and $\bm{\Sigma}_{12} \in \mathbb{C}^{(M-2) \times 2}$ and $\bm{\Sigma}_{22} \in \mathbb{C}^{2 \times (M-2)}$ are zero matrices. The beamforming matrix $\mathbf{A}$ can be structured as
\begin{align}\label{A1}
\mathbf{A} &= \mathbf{V}_{21}\mathbf{\Omega}\mathbf{U}_{11}^H \\
&= \mathbf{V}_{21}\bm{\Lambda}_2\mathbf{U}_2^H\mathbf{V}_1\bm{\Lambda}_1\mathbf{U}_{11}^H, \nonumber
\end{align}
where matrices $\mathbf{\Omega} \in \mathbb{C}^{2 \times 2}$, $\bm{\Lambda}_1 \in \mathbb{C}^{2 \times 2}\succeq \bm{0}$ and $\bm{\Lambda}_2 \in \mathbb{C}^{2 \times 2}\succeq \bm{0}$ are diagonal matrices. $\mathbf{V}_1\bm{\Lambda}_1\mathbf{U}_{11}^H$ denotes receive beamforming, $\mathbf{V}_{21}\bm{\Lambda}_2\mathbf{U}_2^H$ represents transmit beamforming. While $\bm{\Lambda}_1$ and $\bm{\Lambda}_2$ are unknown, which are determined by searching more than four variables satisfying $\bm{\Lambda}_1 \succeq \bm{0}$ and $\bm{\Lambda}_1 \succeq \bm{0}$. Here, ONS method is used to solve $\bm{\Lambda}_1$ and $\bm{\Lambda}_2$ by selecting $\bm{\Lambda}_1=\bm{\Lambda}_2=\sqrt{\rho}\mathbf{I}_2$ \cite{Lyc2008}. \textcolor{blue}{Based on the above derivations,} the beamforming matrix $\mathbf{A}$ at AF relay can be obtained as
\vspace{-0.7em}
\begin{align}\label{A2}
\mathbf{A} &= \rho\mathbf{V}_{21}\mathbf{U}_2^H\mathbf{V}_1\mathbf{U}_{11}^H \\
&= \rho\mathbf{\Upsilon}, \nonumber
\end{align}
where $\mathbf{\Upsilon}=\mathbf{V}_{21}\mathbf{U}_2^H\mathbf{V}_1\mathbf{U}_{11}^H$ and $\rho$ is need to meet transmit power constraint (\ref{op9_6}) with equality. Particularly, $\rho$ is chosen as (\ref{rho}), as shown at the top of next page.
 \begin{figure*}[ht] %hb´ú±í·ÅÔÚÎÄÕµײ¿£¬%htΪ·ÅÔÚÎÄÕ¶¥²¿
	\centering
	\begin{equation}\label{rho}
		\rho =  \sqrt{\frac{P_r}{ \text{tr}\{\bar{\bm{\Theta}}_1(P_1 \mathbf{H}_1^H \mathbf{\Upsilon}^H \mathbf{\Upsilon}\mathbf{H}_1 + P_2 \mathbf{H}_2^H \mathbf{\Upsilon}^H \mathbf{\Upsilon}\mathbf{H}_2)\} +\sigma_r^2 \|\mathbf{\Upsilon}\|_F^2}}.
	\end{equation}
	\hrulefill
\end{figure*}

%\begin{align}\label{rho}
%& \rho =  \\
%& \sqrt{\frac{P_r}{ \text{tr}\{\bar{\bm{\Theta}}_1(P_1 \mathbf{H}_1^H \mathbf{\Upsilon}^H \mathbf{\Upsilon}\mathbf{H}_1 + P_2 \mathbf{H}_2^H \mathbf{\Upsilon}^H \mathbf{\Upsilon}\mathbf{H}_2)\} +\sigma_r^2 \|\mathbf{\Upsilon}\|_F^2}}. \nonumber
%\end{align}

\subsection{Problem Reformulation}
After that, problem (\ref{op3}) can be further translated into the following SDP problem
\begin{subequations}\label{op9}
\begin{align}
&\max_{t, \mathbf{A}, \bar{\bm{\Theta}}_1, \bar{\bm{\Theta}}_2} ~~t  \\
&\text{s.t.} \bar{\bm{\Theta}}_1(n, n)=1, \bar{\bm{\Theta}}_2(n, n)=1, \forall n=1,\cdots,N+1, \label{op9_2}\\
&~ \bar{\bm{\Theta}}_1\succeq \mathbf{0},~ \bar{\bm{\Theta}}_2\succeq \mathbf{0},~ \text{rank}(\bar{\bm{\Theta}}_1)=1,~ \text{rank}(\bar{\bm{\Theta}}_2)=1, \label{op9_3}\\
&~ 2^{2t} \leq 1+ \frac{P_1 \text{tr}(\bar{\bm{\Theta}}_1 \mathbf{H}_1^H \mathbf{A}^H \mathbf{H}_2 \bar{\bm{\Theta}}_2 \mathbf{H}_2^H \mathbf{A} \mathbf{H}_1 ) }{\text{tr}(\bar{\bm{\Theta}}_2\mathbf{D}_{12})}, \label{op9_4}\\
&~ 2^{2t} \leq 1+ \frac{P_2 \text{tr}(\bar{\bm{\Theta}}_1 \mathbf{H}_2^H \mathbf{A}^H \mathbf{H}_1 \bar{\bm{\Theta}}_2 \mathbf{H}_1^H \mathbf{A} \mathbf{H}_2 ) }{\text{tr}(\bar{\bm{\Theta}}_2\mathbf{D}_{21})}, \label{op9_5}\\
&~ \text{tr}\{\bar{\bm{\Theta}}_1(P_1 \mathbf{H}_1^H \mathbf{A}^H \mathbf{A}\mathbf{H}_1 + P_2 \mathbf{H}_2^H \mathbf{A}^H \mathbf{A}\mathbf{H}_2)\} \nonumber \\
&~ +\sigma_r^2 \|\mathbf{A}\|_F^2 \leq  P_r, \label{op9_6}
\end{align}
\end{subequations}
where $\bar{\bm{\Theta}}_1=\bar{\bm{\theta}}_1\bar{\bm{\theta}}_1^H$ and $\bar{\bm{\Theta}}_2=\bar{\bm{\theta}}_2\bar{\bm{\theta}}_2^H$.
Since $\mathbf{A}$, $\bar{\bm{\Theta}}_1$, $\bar{\bm{\Theta}}_2$ are coupled each other and there are two rank-one constraints, which results in a non-convex problem, and directly solving such a non-convex problem is difficult. Meanwhile, given that the closed-form expression of $\mathbf{A}$ in section \ref{section_IV_A} has been obtained, the above mentioned problem (\ref{op9}) can be decomposed into two subproblems associated to $\bar{\bm{\Theta}}_1$ and $\bar{\bm{\Theta}}_2$. The following are the optimization details for obtaining rank-one $\bar{\bm{\Theta}}_1$ and $\bar{\bm{\Theta}}_2$.

\subsection{Optimization of $\bar{\bm{\Theta}}_1$}
Fixing $\mathbf{A}$ and $\bar{\bm{\Theta}}_2$, the optimization problem (\ref{op9}) can be further translated into
\begin{subequations}\label{op10}
\begin{align}
&\max_{t, \bar{\bm{\Theta}}_1} ~~~~~  t  \\
&~~\text{s.t.}~~~~~~ \bar{\bm{\Theta}}_1(n, n)=1,~ \forall n=1,\cdots,N+1, \\
&~~~~~~~~~~~~~ \bar{\bm{\Theta}}_1\succeq \mathbf{0},~ \text{rank}(\bar{\bm{\Theta}}_1)=1, \\
&~~~~~~~~~~~~~ \text{(\ref{op9_4})},~ \text{(\ref{op9_5})},~ \text{(\ref{op9_6})}.
\end{align}
\end{subequations}
The non-convex constraint $\text{rank}(\bar{\bm{\Theta}}_1)=1$ leads the above problem to be non-convex, \textcolor{blue}{so we still cannot directly address it.} With the aim of converting the above problem to a convex problem, we perform the following equivalent operation on the constraint $\text{rank}(\bar{\bm{\Theta}}_1)=1$
\begin{equation}\label{operation1}
\text{tr}(\bar{\bm{\Theta}}_1)-\lambda_{max}(\bar{\bm{\Theta}}_1) \leq 0,
\end{equation}
where $\lambda_{max}(\bar{\bm{\Theta}}_1)$ is the maximum eigenvalue of $\bar{\bm{\Theta}}_1$. Due to the constraint $\bar{\bm{\Theta}}_1\succeq \mathbf{0}$, it is implied that $\text{tr}(\bar{\bm{\Theta}}_1)-\lambda_{max}(\bar{\bm{\Theta}}_1) \geq 0$.
In order to better address the problem of rank-one constraint, we adopt a penalty method to recover rank-one $\bar{\bm{\Theta}}_1$. Firstly, a slack variable $\xi_1 \geq 0$ is introduced to expand the size of the feasible solution $\bar{\bm{\Theta}}_1$ in constraint (\ref{operation1}). Then another relaxation variable $\mu_1 >$ 0, namely penalty parameter, is introduced to the objective function. After that, problem (\ref{op10}) can be further equivalently reformulated as follows
\begin{subequations}\label{op11}
\begin{align}
&\max_{t, \xi_1, \bar{\bm{\Theta}}_1} ~~~  t-\mu_1\xi_1  \\
&~~ \text{s.t.}~~~~ \bar{\bm{\Theta}}_1(n, n)=1,~ \forall n=1,\cdots,N+1, \\
&~ \bar{\bm{\Theta}}_1\succeq \mathbf{0},~\text{tr}(\bar{\bm{\Theta}}_1)-\lambda_{max}(\bar{\bm{\Theta}}_1) \leq \xi_1,~ \xi_1 \geq 0,\\
&~ \text{(\ref{op9_4})},~ \text{(\ref{op9_5})},~ \text{(\ref{op9_6})}.
\end{align}
\end{subequations}
For any $\mu_1 > \mu_1^0$, problem (\ref{op10}) and problem (\ref{op11}) are equivalent, so that the two problems share the same solution. In other words, the penalty optimization solution of problem (\ref{op11}) is also available for problem (\ref{op10}).
Due to the fact that convex function $\lambda_{max}(\bar{\bm{\Theta}}_1)$ is not differentiable, its sub-gradient is written as $\bar{\bm{\theta}}_{max}^1(\bar{\bm{\theta}}_{max}^1)^H$, where $\bar{\bm{\theta}}_{max}^1$ is the eigenvector corresponding to $\lambda_{max}(\bar{\bm{\Theta}}_1)$. Therefore, the first-order approximation of $\lambda_{max}(\bar{\bm{\Theta}}_1)$ is denoted as
\begin{align}\label{first_order1}
\lambda_{max}(\bar{\bm{\Theta}}_1) \geq &\lambda_{max}(\widetilde{\bm{\Theta}}_1) \\
& +\text{tr}(\widetilde{\bm{\theta}}_{max}^1(\widetilde{\bm{\theta}}_{max}^1)^H(\bar{\bm{\Theta}}_1-\widetilde{\bm{\Theta}}_1)),     \nonumber
\end{align}
where $\lambda_{max}(\widetilde{\bm{\Theta}}_1)$ is the maximum eigenvalue of the feasible solution $\widetilde{\bm{\Theta}}_1$, $\widetilde{\bm{\theta}}_{max}^1$ is the eigenvector corresponding to $\lambda_{max}(\widetilde{\bm{\Theta}}_1)$.
Placing the low bound of $\lambda_{max}(\bar{\bm{\Theta}}_1)$ in problem (\ref{op11}), we have
\begin{subequations}\label{Theta_line1}
\begin{align}
&\max_{t, \xi_1, \bar{\bm{\Theta}}_1} ~~~  t-\mu_1\xi_1  \\
&~~~ \text{s.t.}~~ \bar{\bm{\Theta}}_1(n, n)=1,~ \forall n=1,\cdots,N+1, \\
&~~~~ \bar{\bm{\Theta}}_1\succeq \mathbf{0},~ \text{tr}(\bar{\bm{\Theta}}_1)-\lambda_{max}(\widetilde{\bm{\Theta}}_1)-\text{tr}(\widetilde{\bm{\theta}}_{max}^1(\widetilde{\bm{\theta}}_{max}^1)^H \nonumber\\
&~~~~~~~~~~~~~~~~~~~~~~~~~~ \bullet(\bar{\bm{\Theta}}_1-\widetilde{\bm{\Theta}}_1))  \leq \xi_1,~ \xi_1 \geq 0,  \\
&~~~~ \text{(\ref{op9_4})},~ \text{(\ref{op9_5})},~ \text{(\ref{op9_6})}.
\end{align}
\end{subequations}
When $\mu_1$ and $\widetilde{\bm{\Theta}}_1$ are known, the above optimization problem can be efficiently solved via CVX tool, while the rank-one solution $\bar{\bm{\Theta}}_1$ is obtained.

\subsection{Optimization of $\bar{\bm{\Theta}}_2$ }
When $\mathbf{A}$ and $\bar{\bm{\Theta}}_1$ are fixed, the optimization problem (\ref{op9}) can be reduced to as follows
\begin{subequations}\label{op12}
\begin{align}
&\max_{t,\bar{\bm{\Theta}}_2}~~  t  \\
&~~\text{s.t.}~~ \bar{\bm{\Theta}}_2(n, n)=1,~ \forall n=1,\cdots,N+1, \label{op12_1} \\
&~~~~~~~~~ \bar{\bm{\Theta}}_2\succeq \mathbf{0},~ \text{rank}(\bar{\bm{\Theta}}_2)=1, \label{op12_2} \\
&~~~~~~ 2^{2t} \leq 1+ \frac{\beta_1 P \text{tr}(\bar{\bm{\Theta}}_2 \mathbf{H}_2^H \mathbf{A} \mathbf{H}_1 \bar{\bm{\Theta}}_1 \mathbf{H}_1^H \mathbf{A}^H \mathbf{H}_2 )}{\text{tr}(\bar{\bm{\Theta}}_2\mathbf{C}_{12})}, \label{op12_3} \\
&~~~~~~ 2^{2t} \leq 1+ \frac{\beta_2 P \text{tr}(\bar{\bm{\Theta}}_2 \mathbf{H}_1^H \mathbf{A} \mathbf{H}_2 \bar{\bm{\Theta}}_1 \mathbf{H}_2^H \mathbf{A}^H \mathbf{H}_1 )}{\text{tr}(\bar{\bm{\Theta}}_2\mathbf{C}_{21})}. \label{op12_4}
\end{align}
\end{subequations}
To transform the above problem into a easily solvable form, let us define two slack variables as follow
\begin{align}
&\eta_{12}^k=\frac{\sqrt{\beta_1 P \text{tr}(\bar{\bm{\Theta}}_2^k \mathbf{H}_2^H \mathbf{A} \mathbf{H}_1 \bar{\bm{\Theta}}_1 \mathbf{H}_1^H \mathbf{A}^H \mathbf{H}_2 )}}{\text{tr}(\bar{\bm{\Theta}}_2^k\mathbf{C}_{12})}, \\
&\eta_{21}^k=\frac{\sqrt{\beta_2 P \text{tr}(\bar{\bm{\Theta}}_2^k \mathbf{H}_1^H \mathbf{A} \mathbf{H}_2 \bar{\bm{\Theta}}_1 \mathbf{H}_2^H \mathbf{A}^H \mathbf{H}_1 )}}{\text{tr}(\bar{\bm{\Theta}}_2^k\mathbf{C}_{21})}.
\end{align}
In line with GFP algorithm \cite{Skm2018}, constraint (\ref{op12_3}) and (\ref{op12_4}) can be respectively changed to the following two convex constraints
\begin{align}
2^{2t} \leq &1+ 2\eta_{12}^k\sqrt{\beta_1 P \text{tr}(\bar{\bm{\Theta}}_2 \mathbf{H}_2^H \mathbf{A} \mathbf{H}_1 \bar{\bm{\Theta}}_1 \mathbf{H}_1^H \mathbf{A}^H \mathbf{H}_2 )} \nonumber\\
&-(\eta_{12}^k)^2\text{tr}(\bar{\bm{\Theta}}_2\mathbf{C}_{12}), \label{op12_3_1} \\
2^{2t} \leq &1+ 2\eta_{21}^k\sqrt{\beta_2 P \text{tr}(\bar{\bm{\Theta}}_2 \mathbf{H}_1^H \mathbf{A} \mathbf{H}_2 \bar{\bm{\Theta}}_1 \mathbf{H}_2^H \mathbf{A}^H \mathbf{H}_1 )} \nonumber\\
&-(\eta_{21}^k)^2\text{tr}(\bar{\bm{\Theta}}_2\mathbf{C}_{21}). \label{op12_4_1}
\end{align}
We bring the above two inequalities in problem (\ref{op12}), which yields
\begin{subequations}\label{op13}
\begin{align}
&\max_{t,\bar{\bm{\Theta}}_2}~~~  t  \\
&~~\text{s.t.}~~~~ \bar{\bm{\Theta}}_2(n, n)=1,~ \forall n=1,\cdots,N+1, \\
&~~~~~~~~~~~ \bar{\bm{\Theta}}_2\succeq \mathbf{0},~ \text{rank}(\bar{\bm{\Theta}}_2)=1, \\
&~~~~~~~~~~~ \text{(\ref{op12_3_1})},~ \text{(\ref{op12_4_1})}.
\end{align}
\end{subequations}
For the non-convex constraint $\text{rank}(\bar{\bm{\Theta}}_2)=1$, we adopt the same penalty method to process it. For brief, the related details are omitted.
By using the penalty method, problem (\ref{op13}) is converted to
\begin{subequations}\label{Theta_line2}
\begin{align}
&\max_{t, \xi_2, \bar{\bm{\Theta}}_2}~~~  t-\mu_2\xi_2  \\
&~~~\text{s.t.}~~ \bar{\bm{\Theta}}_2(n, n)=1,~ \forall n=1,\cdots,N+1,  \\
&~~~~ \bar{\bm{\Theta}}_2\succeq \mathbf{0},~ \text{tr}(\bar{\bm{\Theta}}_2)-\lambda_{max}(\widetilde{\bm{\Theta}}_2)-\text{tr}(\widetilde{\bm{\theta}}_{max}^2(\widetilde{\bm{\theta}}_{max}^2)^H  \nonumber\\
&~~~~~~~~~~~~~~~~~~~~~~~~~~ \bullet(\bar{\bm{\Theta}}_2-\widetilde{\bm{\Theta}}_2))  \leq \xi_2,~ \xi_2 \geq 0, \\
&~~~~ \text{(\ref{op12_3_1})},~ \text{(\ref{op12_4_1})},
\end{align}
\end{subequations}
where $\mu_2 >$ 0 is a penalty parameter, $\xi_2 \geq 0$ is a slack variable and $\widetilde{\bm{\theta}}_{max}^2$ is the eigenvector corresponding to the maximum eigenvalue $\lambda_{max}(\widetilde{\bm{\Theta}}_2)$ of the feasible solution $\widetilde{\bm{\Theta}}_2$. In the same manner, problem (\ref{Theta_line2}) can be solved by CVX, thereby solution $\bar{\bm{\Theta}}_2$ satisfying rank-one constraint can be achieved.

\subsection{Overall algorithm}

In the proposed ONS-SDP-PSCA scheme, the expression of beamforming matrix $\mathbf{A}$ can be obtained in closed form by ONS method, and IRS rank-one phase matrices $\bar{\bm{\Theta}}_1$ and $\bar{\bm{\Theta}}_2$ can be achieved by a SDP-PSCA-based penalty method. The detailed iterative process of the proposed ONS-SDP-PSCA scheme is presented in Algorithm 2. It is noted that the two penalty factors $\mu_1$ and $\mu_2$ are gradually increased in each sub-iteration of finding rank-one $\bar{\bm{\Theta}}_1$ and $\bar{\bm{\Theta}}_2$, when the $\zeta_{1max}$ and $\zeta_{2max}$ are reached, $\xi_1$ and $\xi_2$ are very small, which are considered as 0.

According to (\ref{A2}) and (\ref{rho}), the complexity of $\mathbf{A}$ is denoted as $2(N+1)^3+4M^2N+4MN^2+9M^2-N^2+6MN+14M+3N+3$ FLOPs. In problem (\ref{Theta_line1}), there exist $n_{\bar{\bm{\Theta}}_1}=(N+1)^2+2$ variables, $N+6$ linear constraints, one linear matrix inequality (LMI) constraint of size $N+1$, its complexity is $n_{\bar{\bm{\Theta}}_1}\sqrt{2N+7}[(N+1)^3+n_{\bar{\bm{\Theta}}_1}((N+1)^2+N+6)+ n_{\bar{\bm{\Theta}}_1}^2+N+6  ]$. Similarly, problem (\ref{Theta_line2}) has $N+5$ linear constraints, one linear matrix inequality (LMI) constraint of size $N+1$. The corresponding complexity is $n_{\bar{\bm{\Theta}}_2}\sqrt{2N+6}[(N+1)^3+n_{\bar{\bm{\Theta}}_2}((N+1)^2+N+5)+ n_{\bar{\bm{\Theta}}_2}^2+N+5]$, where $n_{\bar{\bm{\Theta}}_2}=(N+1)^2+2$ is the number of variables. The total complexity of ONS-SDP-PSCA algorithm is calculated as
\begin{algorithm}
	\caption{Proposed ONS-SDP-PSCA algorithm}
	\begin{algorithmic}[1]
		\STATE Given $\mathbf{A}^0$, $\bar{\bm{\Theta}}_1^0$ and $\bar{\bm{\Theta}}_2^0$, $R^0$ can be computed.
		\STATE Set the convergence error $\delta$ and initialize iteration number $k=k_1=k_2=0$.
		\REPEAT
		\STATE With ($\bar{\bm{\Theta}}_1^{k}$, $\bar{\bm{\Theta}}_2^{k}$), calculate $\mathbf{A}^{k+1}$ with (\ref{A2}) and (\ref{rho}).
		\STATE With ($\mathbf{A}^{k+1}$, $\bar{\bm{\Theta}}_2^{k}$), initialize $\mu_1^0$ and $\widetilde{\bm{\Theta}}_1^0$, set $\zeta_1 > 0$ and $\zeta_{1max}$.
		\REPEAT
		\STATE With ($\mu_1^{k_1}$, $\widetilde{\bm{\Theta}}_1^{k_1}$), obtain ($\xi_1^{k_1+1}$, $\bar{\bm{\Theta}}_1^{k_1+1}$) by solving problem (\ref{Theta_line1}).
		\STATE Update $\mu_1^{k_1+1}=\text{min}\{\zeta_1\mu_1^{k_1}, \zeta_{1max}\}$ and $\widetilde{\bm{\Theta}}_1^{k_1+1}=\bar{\bm{\Theta}}_1^{k_1+1}$.
		\STATE $k_1 = k_1 + 1$.
		\UNTIL (\ref{Theta_line1}) converges, and set $\bar{\bm{\Theta}}_1^{k+1}=\bar{\bm{\Theta}}_1^{k_1+1}$.
		\STATE With ($\mathbf{A}^{k+1}$, $\bar{\bm{\Theta}}_1^{k+1}$), initialize $\mu_2^0$ and $\widetilde{\bm{\Theta}}_2^0$, set $\zeta_2 > 0$ and $\zeta_{2max}$.
		\REPEAT
		\STATE With ($\mu_2^{k_2}$, $\widetilde{\bm{\Theta}}_2^{k_2}$), obtain ($\xi_2^{k_2+1}$, $\bar{\bm{\Theta}}_2^{k_2+1}$) by solving problem (\ref{Theta_line2}).
		\STATE Update $\mu_2^{k_2+1}=\text{min}\{\zeta_2\mu_2^{k_2}, \zeta_{2max}\}$ and $\widetilde{\bm{\Theta}}_2^{k_2+1}=\bar{\bm{\Theta}}_2^{k_2+1}$.
		\STATE $k_2 = k_2 + 1$.
		\UNTIL (\ref{Theta_line2}) converges, and set $\bar{\bm{\Theta}}_2^{k+1}=\bar{\bm{\Theta}}_2^{k_2+1}$.
		\STATE Calculate $R^{k+1}$ with ($\mathbf{A}^{k+1}$, $\bar{\bm{\Theta}}_1^{k+1}$, $\bar{\bm{\Theta}}_2^{k+1}$).
		\STATE $k=k+1$.
		\UNTIL $|R^{k+1} - R^k| \leq \delta$.
		%\STATE $|R^{k+1} - R^k| \leq \delta$.
	\end{algorithmic}
\end{algorithm}
\vspace{-1.0em}
\begin{align}
&\mathcal{O}\{D_2[
2(N+1)^3+4M^2N+4MN^2+9M^2 \\
&-N^2+6MN+14M+3N+3+n_{\bar{\bm{\Theta}}_1}\sqrt{2N+7} \nonumber\\
&\bullet((N+1)^3+n_{\bar{\bm{\Theta}}_1}((N+1)^2+N+6)+ n_{\bar{\bm{\Theta}}_1}^2 \nonumber\\
&+N+6)+n_{\bar{\bm{\Theta}}_2}\sqrt{2N+6}((N+1)^3+n_{\bar{\bm{\Theta}}_2} \nonumber\\
&\bullet((N+1)^2+N+5)+ n_{\bar{\bm{\Theta}}_2}^2+N+5)
]\text{ln}(1/\varepsilon)\} \nonumber
\end{align}
FLOPs, where $D_2$ is the iterative number in Algorithm 2.
%The highest order of computational complexity is $M^2$ and $N^{6.5}$ FLOPs.

\section{Numerical Results Analysis}\label{Results}
To validate the convergence and rate performance of the proposed LC-ZF-SCA and ONS-SDP-PSCA schemes in this section, some numerical simulation results are presented.
Assuming the coordinates of $\text{S}_1$, $\text{S}_2$, IRS (or UAV) and AF relay are (0, 0, 0), (0, 120m, 0), ($-$10m, 60m, 20m) and (10m, 60m, 10m) in three-dimensional (3D) space, and the path loss at distance $d$ is computed by $PL(d)=PL_0-10{\alpha}\text{log}_{10}(\frac{d}{d_0})$. $PL_0$ is the reference path loss at $d_0=1$m, and is generally set as $-$30dB. Besides, $\alpha$ is the path attenuation index of the channel link between transceivers. In this paper, $\alpha_{1i}$, $\alpha_{1r}$, $\alpha_{2i}$, $\alpha_{2r}$ and $\alpha_{ir}$ respectively denote the path attenuation indexes from $\text{S}_1$ to IRS, from $\text{S}_1$ to AF relay, from $\text{S}_2$ to IRS, from $\text{S}_2$ to AF relay and from IRS to AF relay.
The related parameters are chosen as follow: $\alpha_{1i}=\alpha_{2i}=\alpha_{ir}= $2.0, $\alpha_{1r}=\alpha_{2r}= $3.6, $\sigma_1^2= \sigma_2^2= \sigma_r^2= \sigma^2= -$90dBm, and $P_1=P_2=P_r=\frac{1}{3}P$, where $P$ is the total transmit power of the IRS-assisted two-way AF relay network.

\begin{table*}[htbp]
	\centering
	\caption{ \textcolor{blue}{Complexity analysis of the proposed methods}}\label{Table}
	\setlength{\tabcolsep}{3pt}
	\begin{tabular}{|m{3.0cm}<{\centering}|m{13cm}<{\centering}|}
		\hline
		\textcolor{blue}{Methods} & \textcolor{blue}{Complexity ($\varepsilon=0.1$)} \\
		\hline
		\textcolor{blue}{LC-ZF-SCA}  & \textcolor{blue}{$\mathcal{O}\{D_1[M^3+11M^2+10MN+7M+6+n_{\bar{\bm{\theta}}_1}\sqrt{7}((N+1)^2+N^2+n_{\bar{\bm{\theta}}_1}^2+3n_{\bar{\bm{\theta}}_1}+3)
		+n_{\bar{\bm{\theta}}_2}\sqrt{5}(N^2+n_{\bar{\bm{\theta}}_2}^2+3n_{\bar{\bm{\theta}}_2}+3)
		]\}$} \\
		\hline
		\textcolor{blue}{ONS-SDP-PSCA}  & \textcolor{blue}{$\mathcal{O}\{D_2[
		2(N+1)^3+4M^2N+4MN^2+9M^2-N^2+6MN+14M+3N+3+n_{\bar{\bm{\Theta}}_1}\sqrt{2N+7}((N+1)^3+n_{\bar{\bm{\Theta}}_1}((N+1)^2+N+6)+ n_{\bar{\bm{\Theta}}_1}^2+N+6)+n_{\bar{\bm{\Theta}}_2}\sqrt{2N+6}((N+1)^3+n_{\bar{\bm{\Theta}}_2}((N+1)^2+N+5)+ n_{\bar{\bm{\Theta}}_2}^2+N+5)
		]\}$} \\
		\hline
	\end{tabular}
\end{table*}

%
%\begin{table}[h]
%\centering
%\caption{ʾÀý±í¸ñ}
%\begin{tabular}{|l|c|r|}
%\hline
%×ó¶ÔÆë & ¾ÓÖжÔÆë & ÓÒ¶ÔÆë \\ \hline
%Êý¾Ý1 & Êý¾Ý2 & Êý¾Ý3 \\
%Êý¾Ý4 & Êý¾Ý5 & Êý¾Ý6 \\ \hline
%\end{tabular}
%\end{table}

In order to better analyze the rate performance of the proposed two methods, the following two cases are regarded as the benchmark schemes.

(1) \textbf{IRS-assisted two-way AF relay network with random phase}: With $\textbf{A}$ optimized, the phase of each IRS unit is selected randomly from the phase interval (0, $2\pi$].

(2) \textbf{Only AF relay}: A wireless network aided by an two-way AF relay is considered, while $\textbf{A}$ can be obtained by ONS method.

\textcolor{blue}{Table 1 presents the the complexity of the proposed LC-ZF-SCA and ONS-SDP-PSCA schemes directly. Clearly, the highest order of LC-ZF-SCA is $\mathcal{O}(N^3+M^3)$ FLOPs while that of ONS-SDP-PSCA is $\mathcal{O}(N^{6.5}+M^2)$ FLOPs.}

\begin{figure}[htb]
\centering
\includegraphics[width=0.500\textwidth]{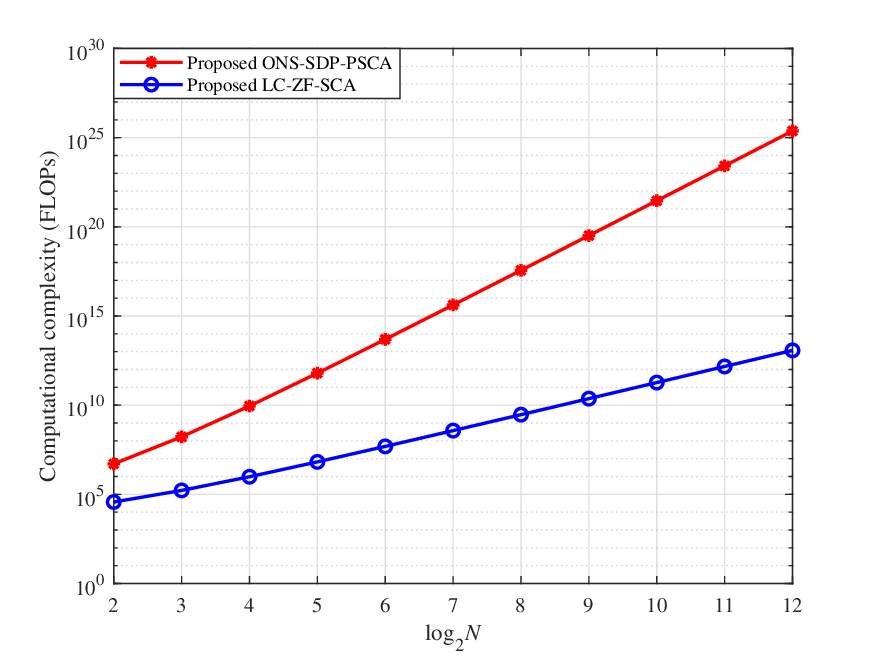}
\centering
\caption{Complexity versus the number $N$ of IRS units given $(M, D1, D2, \varepsilon)=($2, 6, 6, 0.1$)$.}
\label{Complexity_N}
\end{figure}
\begin{figure}[htb]
\centering
\includegraphics[width=0.500\textwidth]{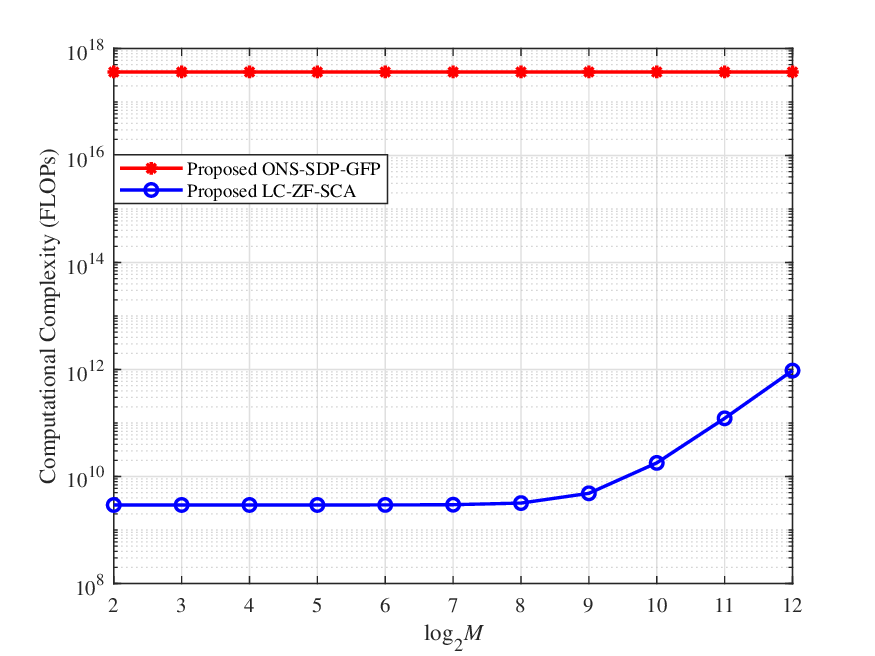}\\
\caption{  \textcolor{blue}{Complexity versus the number $M$ of AF relay antennas given $(N, D1, D2, \varepsilon)=($256, 6, 6, 0.1$)$.}  }
\label{Complexity_M}
\end{figure}
\textcolor{blue}{Fig. \ref{Complexity_N} and Fig. \ref{Complexity_M} show the computational complexity of the proposed two methods versus the number $N$ of IRS units with $(M, D1, D2, \varepsilon)=($2, 6, 6, 0.1$)$ and the number $M$ of AF relay antennas with $(N, D1, D2, \varepsilon)=($256, 6, 6, 0.1$)$. It is obvious that the complexity corresponding to the proposed LC-ZF-SCA and ONS-SDP-PSCA schemes gradually increase as $N$ and $M$ increase. Furthermore, since the optimization variables are matrices, the complexity of ONS-SDP-PSCA method is much higher than that of LC-ZF-SCA method with vector optimization variables.}

\begin{figure}[htb]
	\centering
	\includegraphics[width=0.500\textwidth]{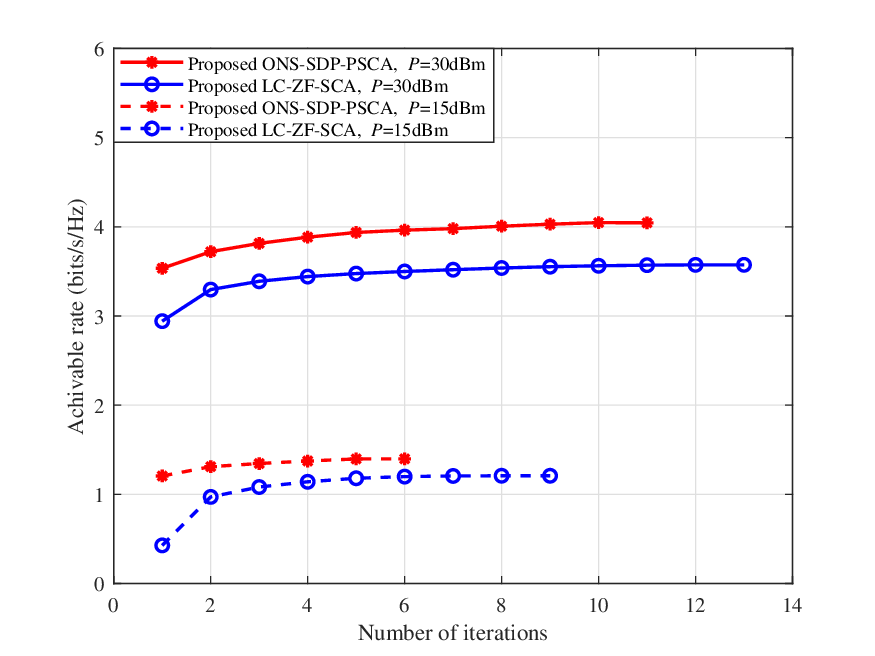}
	\centering
	\caption{Convergence of the proposed two methods given $(M, N)=($2, 128$)$.}
	\label{Iteration}
\end{figure}
Fig. \ref{Iteration} verifies the convergence of the proposed LC-ZF-SCA and ONS-SDP-PSCA methods at $P=$ 15dBm and $P=$ 30dBm, respectively. From Fig. \ref{Iteration}, it is clearly visible that the proposed LC-ZF-SCA and ONS-SDP-PSCA methods can gradually converge to the rate ceil within several iterations for different $P$. For the proposed two schemes, their convergence rates at $P=$ 15dBm are much faster than those at $P=$ 30dBm. Besides, the convergence rate of ONS-SDP-PSCA scheme is faster than that of LC-ZF-SCA method for different $P$. Therefore, we can draw a conclusion that the proposed LC-ZF-SCA and ONS-SDP-PSCA methods are effective and feasible.

\begin{figure}[htb]
\centering
\includegraphics[width=0.500\textwidth]{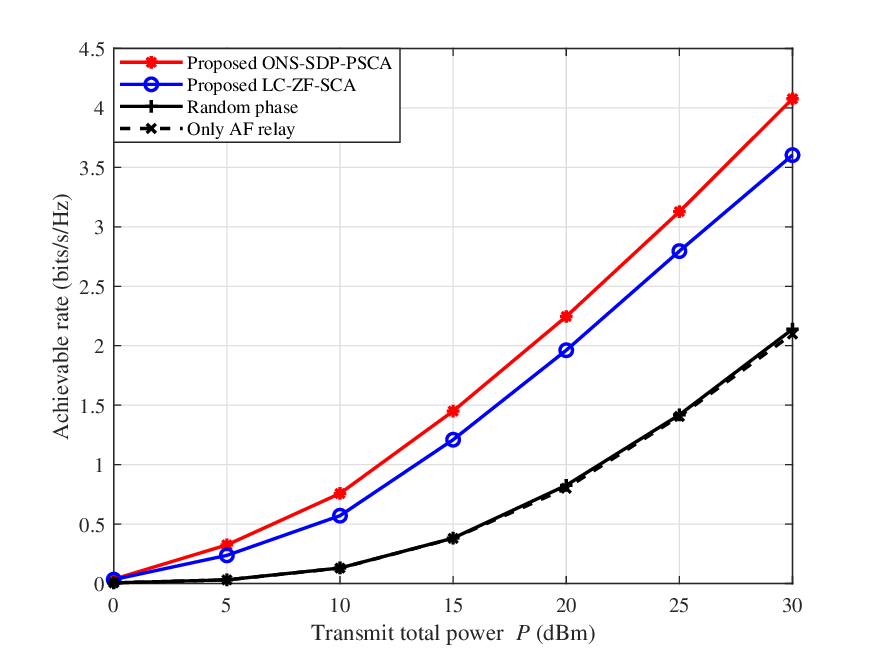}
\centering
\caption{Achievable rate versus total transmit power given $(M, N)=($2, 128$)$.}
\label{Rate_Vs_P}
\end{figure}
Fig. \ref{Rate_Vs_P} presents the achievable rate versus total transmit power $P$ given $(M, N)=($2, 128$)$. As shown in Fig. \ref{Rate_Vs_P}, it is clear that the achievable rates of the proposed two schemes, called LC-ZF-SCA and ONS-SDP-PSCA, increase as total transmit power $P$ increase. In contrast to the two benchmark schemes: random phase and only AF relay, the rate performance enhancement obtained by LC-ZF-SCA and ONS-SDP-PSCA methods are significant in the high $P$ region. Moreover, the proposed ONS-SDP-PSCA method perform better than the proposed LC-ZF-SCA scheme in the all $P$ region. When $P=$ 30dBm, the proposed LC-ZF-SCA and ONS-SDP-PSCA methods can respectively achieve rate performance gains of up to 90.6\% and 68.5\% over those of random phase and only AF relay. Furthermore, the rate performance of ONS-SDP-PSCA method is higher 0.4bits/s/Hz than that of LC-ZF-SCA method.

\begin{figure}[htb]
\centering
\includegraphics[width=0.500\textwidth]{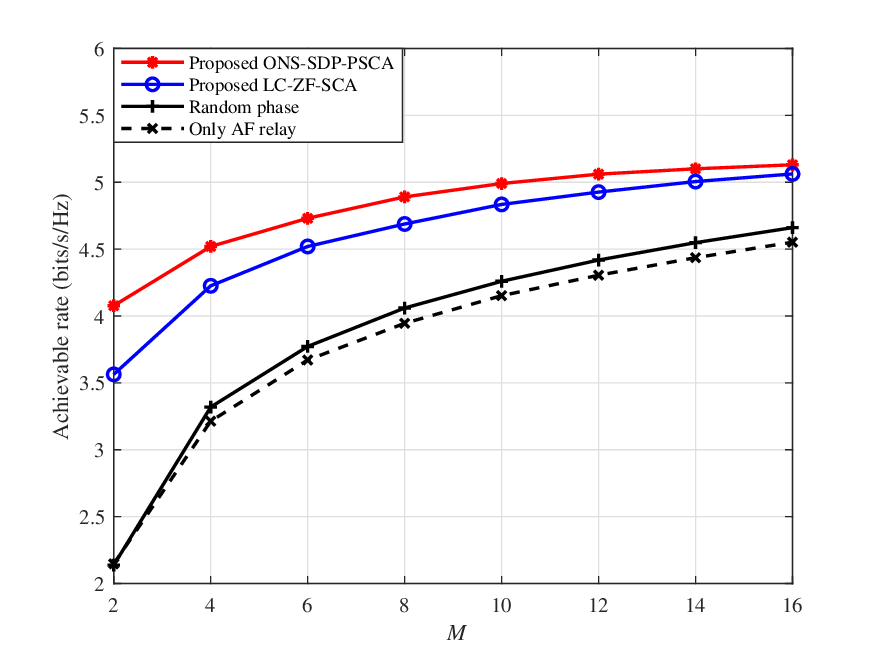}
\centering
\caption{Achievable rate versus the number of antennas at AF relay given $(N, P)=($128, 30dBm$)$.}
\label{Rate_Vs_M}
\end{figure}
Fig. \ref{Rate_Vs_M} shows the achievable rate versus the number $M$ of antennas at AF relay given $(N, P)=($128, 30dBm$)$. As seen in Fig. \ref{Rate_Vs_M}, the two proposed LC-ZF-SCA and ONS-SDP-PSCA methods have higher rate performance than random phase and only AF relay. As $M$ increases, the rate performance of the two proposed methods and the two benchmark schemes increases. Besides that, it can be observed that the decreasing order on rate is ONS-SDP-PSCA, LC-ZF-SCA, random phase and only AF relay.

\begin{figure}[htb]
\centering
\includegraphics[width=0.500\textwidth]{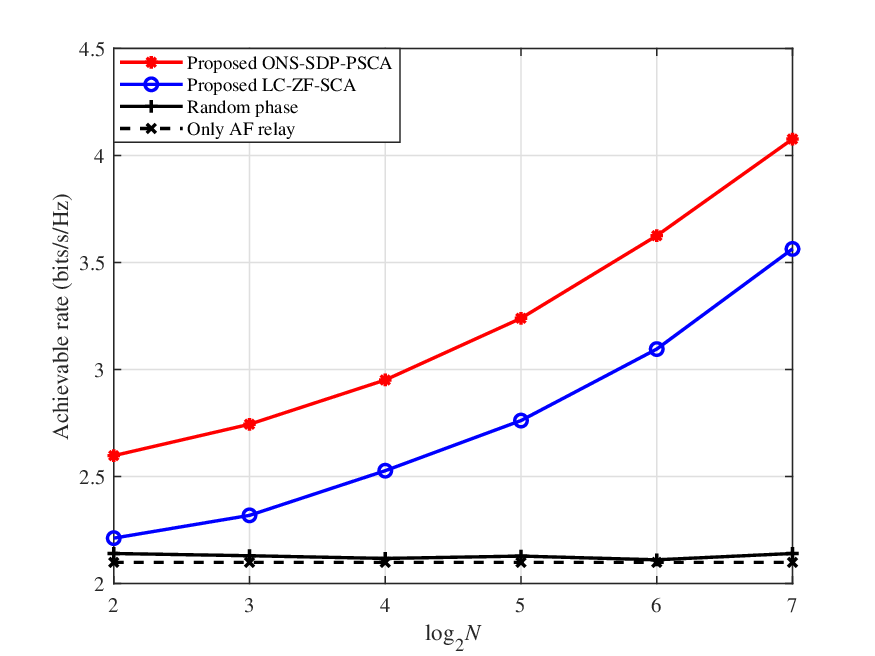}
\centering
\caption{Achievable rate versus the number of IRS units given $(M, P)=($2, 30dBm$)$.}
\label{Rate_Vs_N}
\end{figure}
Fig. \ref{Rate_Vs_N} depicts the achievable rate versus the number $N$ of IRS units for the proposed LC-ZF-SCA and ONS-SDP-PSCA methods with $(M, P)=($2, 30dBm$)$. Apparently, the rate performance gaps between the proposed two methods and the two benchmark schemes gradually widen as $N$ increases, while the rates of the two benchmark schemes remain a small fluctuation. Thus, it is confirmed that optimizing AF beamforming matrix and IRS phase shifts of two time slots is necessary and effective. When $N$ goes to medium and large-scale, the rate performance gaps become especially more evident. Additionally, it can be found that the rate obtained by the proposed LC-ZF-SCA method is lower than that of the proposed ONS-SDP-PSCA method.

\section{Conclusions} \label{Conclusions}

An IRS-and-UAV-aided two-way AF relay network in maritime IoT was discussed in this paper, where two ships were viewed as data centers collecting information from buoys, offshore platforms and sensor nodes. Besides that, the two ships could communicate with the help of an IRS attached to UAV and an AF relay. To solve the problem of maximizing minimum rate, there existed two AI methods called LC-ZF-SCA and ONS-SDP-PSCA were proposed to jointly optimize AF beamforming matrix and IRS phase shifts for rate enhancement.
As shown in simulation results, the proposed LC-ZF-SCA and ONS-SDP-PSCA methods were proved to be convergent and feasible. Compare with an IRS-assisted two-way AF relay network with random phase and only an AF relay network, the achievable rate of the IRS-aided two-way AF relay network could be significantly improved by using the proposed LC-ZF-SCA and ONS-SDP-PSCA methods. For example, at least 68.5\% rate gain could be obtained by the proposed two schemes when $P$=30dBm. Furthermore, the complexity of LC-ZF-SCA method is much lower than that of ONS-SDP-PSCA method at cost of rate performance loss.

\ifCLASSOPTIONcaptionsoff
  \newpage
\fi

\bibliographystyle{IEEEtran}
\bibliography{IEEEfull,reference}

\end{document}